\documentclass[twocolumn,english,aps,superscriptaddress,prb,showpacs]{revtex4-1}
\usepackage{amsmath}
\usepackage{amssymb}
\usepackage{graphicx}
\usepackage[colorlinks,bookmarks=false,citecolor=red,linkcolor=blue,urlcolor=blue]{hyperref}
\usepackage{slashed}
\def\Hhat{\hat{H}}

\def\r{{\boldsymbol r}}
\def\i{{\boldsymbol i}}
\def\l{{\boldsymbol l}}
\def\j{{\boldsymbol j}}
\def\k{{\boldsymbol k}}
\def\K{{\boldsymbol K}}
\def\q{{\boldsymbol q}}

\def\identity{\hat{\mathbf{1}}}
\newcommand{\ve}[1]{\boldsymbol{#1}}

\begin{document}
\title{Kondo breakdown in a spin-1/2 chain of adatoms  on a Dirac semimetal}
\author{Bimla Danu}
\email{bimla.danu@physik.uni-wuerzburg.de}
\affiliation{Institut f\"ur Theoretische Physik und Astrophysik and W\"urzburg-Dresden Cluster of Excellence ct.qmat, Universit\"at W\"urzburg, 97074 W\"urzburg, Germany}
\author{Matthias Vojta}
\email{matthias.vojta@tu-dresden.de}
\affiliation{Institut f\"ur Theoretische Physik and W\"urzburg-Dresden Cluster of Excellence ct.qmat, Technische Universit\"at Dresden, 01062 Dresden, Germany}
\author{Fakher F. Assaad}
\email{fakher.assaad@physik.uni-wuerzburg.de}
\affiliation{Institut f\"ur Theoretische Physik und Astrophysik and W\"urzburg-Dresden Cluster of Excellence ct.qmat, Universit\"at W\"urzburg, 97074 W\"urzburg, Germany}
\author{Tarun Grover}
\email{tagrover@ucsd.edu}
\affiliation{Department of Physics, University of California at San Diego, La Jolla, CA 92093, USA}
\date{\today}
\begin{abstract}
We consider a  spin-1/2 Heisenberg chain  coupled  via a Kondo interaction to  two-dimensional Dirac fermions. The Kondo interaction is irrelevant  at the  \textit{decoupled}  fixed-point, leading to the existence of a Kondo-breakdown phase and a Kondo-breakdown critical point separating such a phase from a heavy Fermi liquid.  We reach this conclusion on the basis of a renormalization group analysis, large-N calculations as well as  extensive auxiliary-field  quantum Monte Carlo simulations. We extract quantities such as the zero-bias tunneling conductance which will be relevant to future experiments involving adatoms on semimetals such as graphene.
\end{abstract}

\maketitle
The antiferromagnetic Kondo coupling,  $J_k$, between a spin-1/2 degree of freedom and a Fermi sea with finite density of states at the Fermi energy is (marginally) relevant: $J_k$ flows to strong coupling and  the impurity is screened. If, in contrast,  the density of states shows a power-law pseudogap behavior, the Kondo coupling is  irrelevant at the decoupled fixed point, and the spin remains unscreened at weak coupling. Since for large Kondo coupling screening is present, a novel Kondo-breakdown quantum critical point emerges~[\onlinecite{Withoff90}, \onlinecite{Fritz04}, \onlinecite{Fritz13}].  The decoupled as well as Kondo-screened phase share the same symmetry properties.

In the context of Kondo lattices, the numbers of both conduction electrons and impurity spins scale with the volume of the system. In the Kondo-screened paramagnetic (i.e. heavy Fermi liquid) phase, the volume enclosed by the Fermi surface (i.e. Luttinger volume) counts both spins and electrons. A Kondo-breakdown transition (equivalently, an orbital-selective Mott transition [\onlinecite{vojta2010osmott}]), which, as above, does not involve  symmetry breaking, implies that the spins drop out from  the Luttinger count. For the case of an odd number of electrons and spins per unit cell, this leads to a violation of the Luttinger sum rule.  Oshikawa's flux-threading argument~[\onlinecite{Oshikawa97}, \onlinecite{Oshikawa2000}] shows that a specific family of the resulting states of matter can be achieved via topological degeneracy in the spin sector~[\onlinecite{Senthil2003}]. Such states, coined fractionalized Fermi liquid (FL$^\ast$) phases, have been realized numerically~[\onlinecite{Hofmann2019}]. Kondo breakdown has also been proposed to understand the phenomenology of heavy-Fermion systems~[\onlinecite{Senthil2003}, \onlinecite{coleman2001}, \onlinecite{Si01}], especially in the context of materials such as  YbRh$_2$Si$_2$ and CeCu${}_{6-x}$Au${}_x$~[\onlinecite{Paschen04}, \onlinecite{Klein08}].

\begin{figure}
\includegraphics[width=0.44\textwidth]{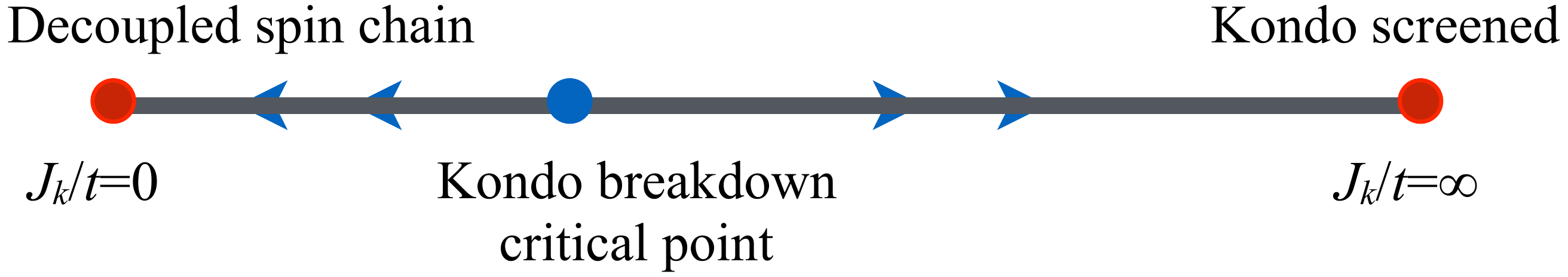}%
\caption{Renormalization group flow of the Kondo coupling, $J_k$,  for  a  spin-1/2 chain on a semimetallic substrate. }
\label{phases_vs_Jk}
\end{figure}

In this article, we consider a situation intermediate between Kondo impurity and Kondo lattice: a one-dimensional (1D) Heisenberg chain which is Kondo-coupled to Dirac electrons. Dimensional analysis shows that, at the decoupled fixed point, the Kondo coupling is irrelevant, thus leading to an RG flow very similar to that of the pseudogap Kondo effect discussed above, see Fig.~\ref{phases_vs_Jk}. The motivation to study  such systems equally stems from scanning tunneling microscopy (STM) experiments of Co adatoms on a Cu$_2$N/Cu(100) surfaces. Here, recent experiments  show an impressive ability to tune the exchange coupling between adatoms as well as the coupling of adatoms to the surface~[\onlinecite{Spinelli2015,Toskovic2016,Choi2017,Moro2019,Jakob2011,Zhou2010,Serrate2010}].  As shown in  Ref.~[\onlinecite{Danu2019}], simple models amenable to  negative-sign-free quantum Monte Carlo (QMC) simulations are able to provide a detailed account of the experiments. Another experimental system that has qualitative resemblance with our setup is Yb${}_2$Pt${}_2$Pb, where neutron scattering indicates the presence of 1D spinons, and apparent absence of Kondo screening, despite the presence of three-dimensional conduction electrons~[\onlinecite{wu2016orbital}, \onlinecite{gannon2019spinon}]. In our study,  we  consider conduction electrons in two dimensions with Dirac spectrum since this choice unambiguously leads to a Kondo-breakdown phase and phase transition, while also allowing RG and large-N calculations and explicit comparison to QMC numerics.

\textit{Model Hamiltionian:} We consider a spin-1/2  Heisenberg chain on a semimetallic substrate:
\begin{eqnarray}
\Hhat &= & -t\sum_{\langle \i,\j\rangle, \sigma}\Big( e^{\frac{2\pi i}{\Phi_0}\int^\j_\i \ve{A}.\ve {d\l}}~\hat{\ve{c}}^{\dagger}_{\i} \hat{\ve{c}}_{\j}+h.c.\Big)\nonumber\\
 & & +\frac{J_k}{2}\sum^L_{\ve{l} =1}  \hat{\ve{c}}^{\dagger}_{\ve{l}}  \ve{\sigma} \hat{\ve{c}}^{}_{\ve{l}} \cdot \hat{\ve{S}}_{\ve{l}}+J_h\sum^{L}_{\ve{l}=1} \hat{\ve{S}}_{\ve{l}} \cdot \hat{\ve{S}}_{\ve{l}+\Delta\ve{l}}.
\label{model_ham}
\end{eqnarray}
Here, $t$ is the hopping parameter of the conduction electrons, the summation $\sum_{\langle \i,\j\rangle}$ runs over a  square lattice and $\hat{\ve{c}}^{\dagger}_{\ve{i}} = \big(\hat{c}^{\dagger}_{\ve{i},\uparrow}, \hat{c}^{\dagger}_{\ve{i},\downarrow}\big) $  is a spinor  where  $ \hat{c}^{\dagger}_{\ve{i},\uparrow (\downarrow)} $ creates an electron at site $\ve{i}$ with $z$-component of spin $1/2$ ($-1/2$). We use the Landau gauge, $\ve{A}=B(-y,0,0)$,  and tune $B$ such that half a flux quantum ($\pi$-flux)  pierces each plaquette. This gauge choice allows for translation symmetry by one  lattice site  in the $x$-direction.   $J_k>0$  is the antiferromagnetic Kondo coupling between  magnetic adatoms and  conduction electrons, $J_h>0$  the Heisenberg coupling between magnetic adatoms, $L$ the length of the Heisenberg chain and linear length of the square conduction electron lattice, and  $\hat{\ve{S}}_{\ve{l}}$ represents the spin-1/2 operators. We use an array of adatoms at interatomic distance $\Delta \ve{l} =  (1,0) $  on  the  substrate and choose periodic boundary conditions along the spin chain and on the substrate to access the thermodynamic limit.

\textit{RG analysis:}   Consider the Hamiltonian in Eq.~(\ref{model_ham}) at $J_k=0$. At low energies, this describes two decoupled conformal field theories (CFT): a (2+1)-D CFT corresponding to Dirac fermions, and a (1+1)-D CFT corresponding to SU(2)$_1$ WZW description of the spin-1/2 Heisenberg chain (we ignore the marginal perturbations that lead to multiplicative logarithmic corrections to the power-law correlations in the chain). The scaling dimension of Dirac fermions  in $d$ space dimensions reads $\Delta_{\psi} = \frac{d}{2}$ and for the spin-1/2 chain, $\Delta_{S} = \frac{1}{2}$.   At this decoupled fixed point, the Kondo coupling  has a scaling dimension $2-2\Delta_\psi-\Delta_S=2-d-\frac{1}{2}=-\frac{1}{2}$  and is thereby irrelevant. On the other hand, in the limit $J_k \rightarrow \infty$ each spin-1/2 degree of freedom  binds in a singlet  with  a conduction electron. This one-dimensional singlet product state, corresponding to the strong-coupling limit of the one-dimensional Kondo lattice model~[\onlinecite{Tsunetsugu1997}], decouples from the conduction electrons, and effectively changes the boundary condition in the $y$-direction from periodic to open.  At large but finite $J_k$, we expect the system to be locally described by a heavy Fermi liquid. Assuming these two regimes are separated by a single phase transition motivates us to find a suitable renormalization group (RG) description of the critical point separating the two regimes. The approach we follow is to consider $(d+1)$-dimensional Dirac fermions coupled to (1+1)-D Heisenberg chain. By power-counting, the Kondo coupling is marginal in $d=3/2$, which allows for an expansion in $\epsilon=d-3/2$, where the physical case of interest corresponds to $d = 2$, i.e., $\epsilon = 1/2$. Perturbing around the $J_k = 0$ fixed point, the RG flow of dimensionless Kondo coupling $j_k = J_k \Lambda^{\epsilon}$ is given by:
\begin{equation}
 \frac{d j_k}{d  \ln \Lambda}  =  \epsilon j_k  - \frac{j_k^2}{2}
 \label{flow_eq_Jk}
\end{equation}
where $\Lambda$ is an ultraviolet cutoff, and we have kept terms to $O(j_k^2)$ (see Sec.~\ref{RG_cal} of  Ref.~[\onlinecite{suppl}] for details). The resulting flow diagram is shown in Fig.~\ref{phases_vs_Jk}  and the  Kondo-breakdown critical fixed point is given by $j^{c}_k = 2 \epsilon$, which yields the correlation length exponent $\nu = 1/\epsilon$. Due to Lorentz invariance, the critical theory will exhibit $\omega/T$ scaling in all observables.

\begin{figure}[h]
\includegraphics[width=0.48\textwidth]{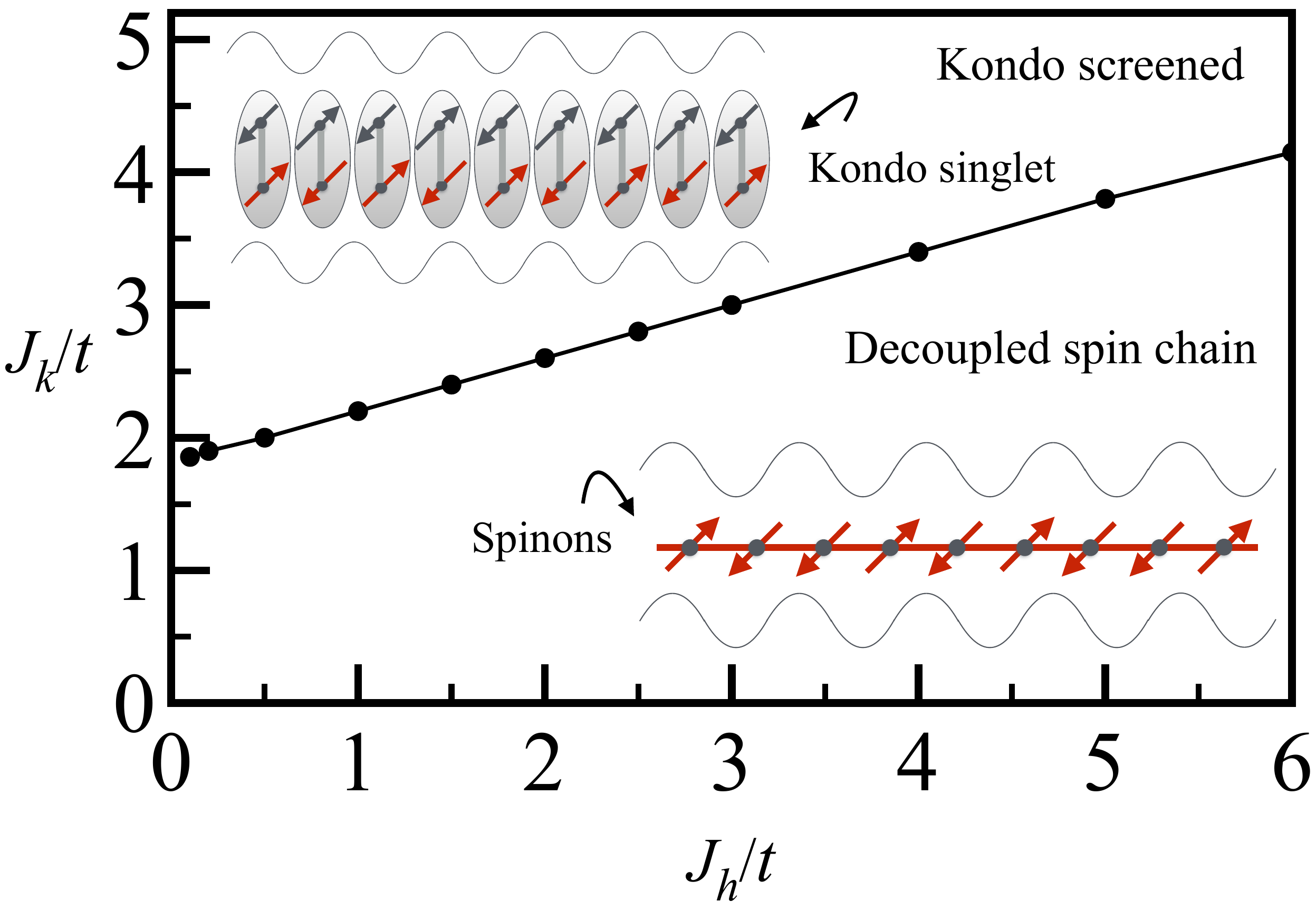}
\caption{The zero-temperature mean-field phase diagram in a parameter space of $J_k/t$ and $J_h/t$. The critical line with symbols separates the two phases.}
\label{meanfld_phase_Jh_Jk}
\end{figure}

\textit{Large-N approximation:}
To formulate the large-N approximation, we  use a  fermion representation of the spin degree of freedom,
$\ve{\hat{S}}_{\ve{l}}=\frac{1}{2} \hat{d}^\dagger_{\ve{l} } \ve{\sigma} \hat{d}_{\ve{l}}$
and impose the constraint $\ve{\hat{d}}^\dagger_{\ve{l}} \ve{\hat{d}}_{\ve{l} }=1$  with
$\ve{\hat{d}}^{\dagger}_{\ve{l}} =\big(\hat{d}^{\dagger}_{\ve{l},\uparrow}, \hat{d}^{\dagger}_{\ve{l},\downarrow}\big)$.
The interaction part of the Hamiltonian can then  be written as:
$-\frac{J_k}{4} \sum_{\ve{l}} \big(\ve{\hat{c}}^{\dagger}_{\ve{l} }\ve{\hat{d}}^{}_{\ve{l} }+h.c.\big)^2-\frac{J_h}{4} \sum_{\ve{l}}\big(\ve{\hat{d}}^{\dagger}_{\ve{l} }\ve{\hat{d}}^{}_{\ve{l}+\Delta \ve{l}}+h.c.\big)^2+\frac{U}{2}\sum_{\ve{l}} \big(\ve{\hat{d}}^{\dagger}_{\ve{l} }\ve{\hat{d}}^{}_{\ve{l} }-1 \big)^2$. We now let the spin-index run from 1 to $N$, and take $N$ to infinity, which allows us to obtain the phase diagram in Fig.~\ref{meanfld_phase_Jh_Jk}  using the saddle-point approximation. The saddle-point variables are determined by:  $V=\sum_{\sigma} \langle \hat{c}^{\dagger}_{l,0,\sigma}  \hat{d}_{l,0,\sigma} \rangle$,  $\chi = \sum_\sigma \langle  \hat{d}^{\dagger}_{l,\sigma} \hat{d}^{}_{l+1,\sigma} \rangle$  and  $\sum_{\sigma} \langle \hat{d}^\dagger_{l,\sigma} \hat{d}_{l,\sigma} \rangle =1$.
The details of the calculations are  presented in Secs.~\ref{saddlepoint_QMC} and~\ref{meanfield_cal} of  Ref.~[\onlinecite{suppl}].  Within this approximation, Kondo breakdown  corresponds to  the solution $V=0$ and $\chi \ne 0$ and Kondo screening to   $V\ne 0$ and $\chi \ne 0$. As apparent, for each value of $J_h$  the mean-field  solution shows  a single transition.    In the limit $J_h =0$, the critical value of $J_k$ corresponds to that  of the single-impurity pseudogap Kondo problem.
Aside from the mean-field order parameters, the  transition  can be detected by  considering the spin-spin correlations along the  chain. In the  decoupled phase spinons are confined to   chain and the spin-spin correlations -- at the mean-field level -- decay as $1/r^2$.   In the Kondo-screened phase, spins hybridize with the Dirac electrons. Since the spin system is sub-extensive, the properties of the Dirac electrons remain unchanged and the spin-spin correlations along the chain inherit the 2D Dirac $1/r^4$   decay (see Fig.~\ref{meanfield_correlation} of Ref.~[\onlinecite{suppl}]).
Introducing particle-hole asymmetry by adding next-nearest hopping (while keeping a half-filled semimetallic state) was found to lead to similar results within large-N~[\onlinecite{suppl}].

\textit{QMC simulations:}  We  have used the Algorithms for Lattice Fermions (ALF) [\onlinecite{ALF_v1}] implementation of the finite-temperature auxiliary-field QMC algorithm~[\onlinecite{Blankenbecler81,White89,Assaad08_rev}].  The  perfect square form of the interaction used to formulate the  large-N calculation complies with the standards of the ALF-library and  the  model can be readily implemented by decoupling  the perfect square terms with a Hubbard Stratonovich transformation. The absence of  negative sign problem follows by first carrying  out a  partial particle-transformation, $\hat{d}^{\dagger}_{ \ve{l}, \uparrow} \rightarrow  e^{i \ve{Q}\cdot \ve{l}} \hat{d}^{}_{\ve{l},\uparrow}$, and  $\hat{c}^{\dagger}_{ \ve{l}, \uparrow} \rightarrow -e^{i \ve{Q}\cdot \ve{l}} \hat{c}^{}_{\ve{l},\uparrow}$,   and then using time reversal symmetry to prove that the eigenvalues of the fermion matrix occur in complex conjugate pairs.  For a given  system of  linear length  $L$, the QMC simulations are performed at an inverse temperature $\beta(=1/k_BT)=L$ and at  a fix $J_h/t=1$.   At $L=20$ we checked  that the the choice $\beta=2L$ shows similar results as $\beta=L$. For the considered periodic boundary conditions, $L=4n+2$ corresponds to open-shell configurations and is known to show less finite-size effects than  $L=4n+4$ sized systems.

\begin{figure}[]
\includegraphics[width=0.49\textwidth]{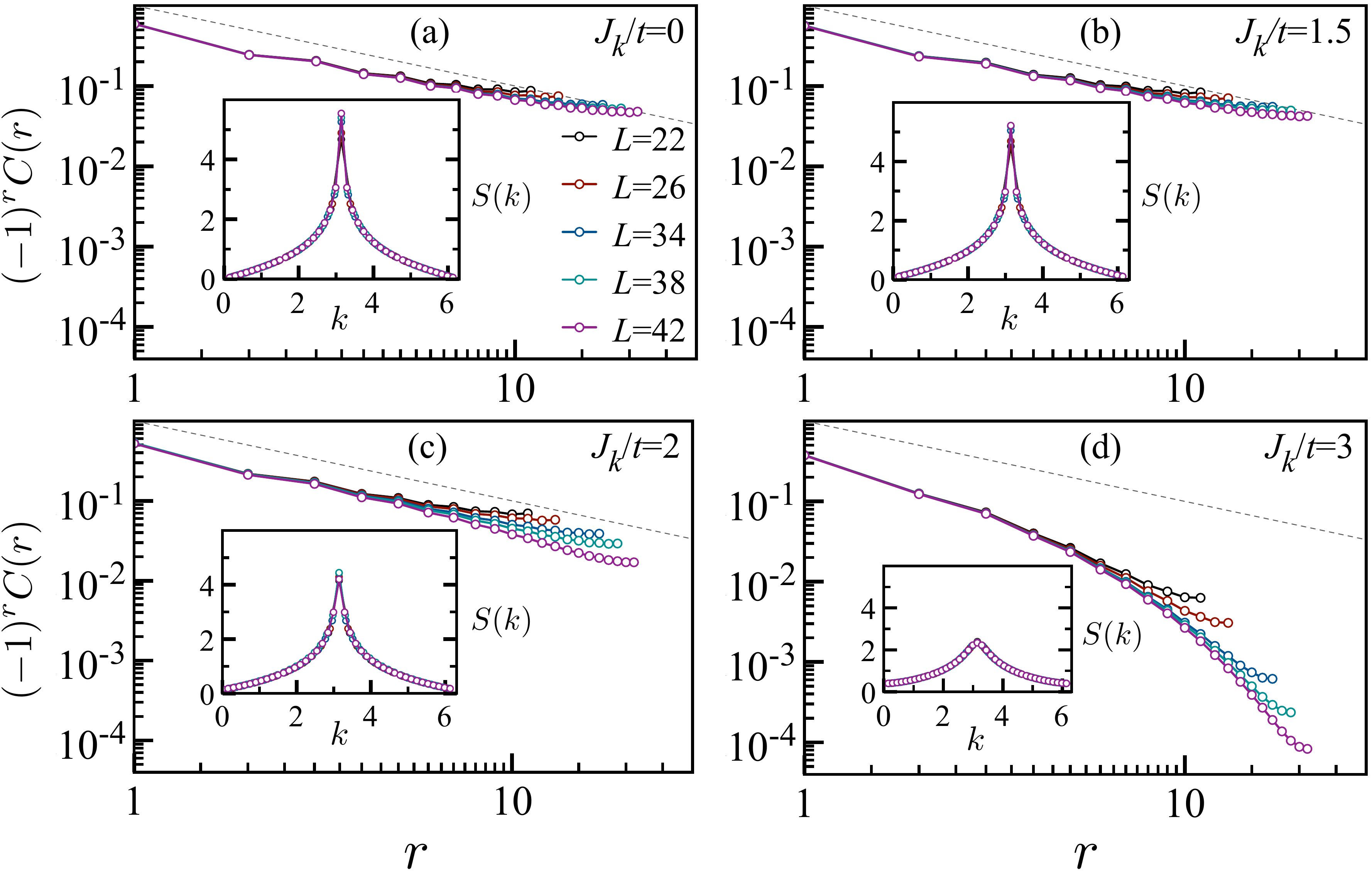}
\caption{Equal-time spin-spin correlation function, $C(r)$, as a function of distance $r$ along the spin chain on a  log-log scale  for  various values of $J_k/t$ at $J_h/t=1$ and $L_x=L_y=L=\beta$. The grey dashed line  corresponds to $1/r$ decay and the corresponding  static spin structure factors $S(k)$ are shown in the insets.}\label{LogSr_vs_logr_4n+2}
\end{figure}
\textit{QMC results:}
Fig.~\ref{LogSr_vs_logr_4n+2}  plots the spin-spin correlations  $C(r) =4\langle{\hat S^z}_{0} {\hat S^z}_ {r}\rangle$ as a function of distance $r$  for various   values of $J_k/t$.  In the limit of vanishing Kondo coupling, our results are consistent with the exact  asymptotic form: $ C(r) \propto(-1)^r\sqrt{\ln r}/r$. The $1/r$  decay of the spin-spin correlations in the Heisenberg model, is tied to  SU(2) spin symmetry.  If the Kondo  coupling is irrelevant, then  we expect $\sum_{\ve{l}}   \hat{\ve{S}}_{\ve{l}} $ to remain a good quantum number  of the low-energy effective theory. Thereby  the asymptotic form of the spin-spin correlations  should equally follow a $(-1)^{r}/r$ form. Remarkably,  the data supports this point of view up to $J_k/t\lesssim2$.   On the other hand, in the Kondo-screened phase for $J_k/t\gtrsim2$, the equal-time correlations decay with a power larger than unity.   In this phase, we expect the spin-spin correlations to inherit the  power-law of the Dirac fermions $\langle { \hat S^{z,c}}_{\ve{l}} {\hat S^{z,c}}_{\ve{l}+\r} \rangle  \propto 1/r^4$.  (see Fig.~\ref{meanfield_correlation} of Ref.~[\onlinecite{suppl}]).
The  insets  of Fig.~\ref{LogSr_vs_logr_4n+2} plot the static spin structure factor $S(k)= \frac{1}{L}\sum_{r} e^{-ik \cdot r}C(r)$  as a function of momentum $k$. Noticeably, both at $J_k=0$ and $J_k/t= 1.5$ we observe  systematic growth of   $S(k)$  at $k = \pi$, reflecting the $(-1)^r/r$ real space decay.  At $J_k/t =2 $ we observe a cusp feature but a  saturation of $S(k = \pi)$  with  system size thus suggesting  a power law  with exponent $ 1 <  K_\sigma  < 2 $.
Finally, in the Kondo-screened phase at $J_k/t=3$,  $S(k)$   converges to a smooth function  implying $K_\sigma > 2$.  A detailed overview of the QMC data is given in Sec.~\ref{QMC_data_details} of Ref.~[\onlinecite{suppl}].

 \begin{figure}[]
\includegraphics[width=0.235\textwidth]{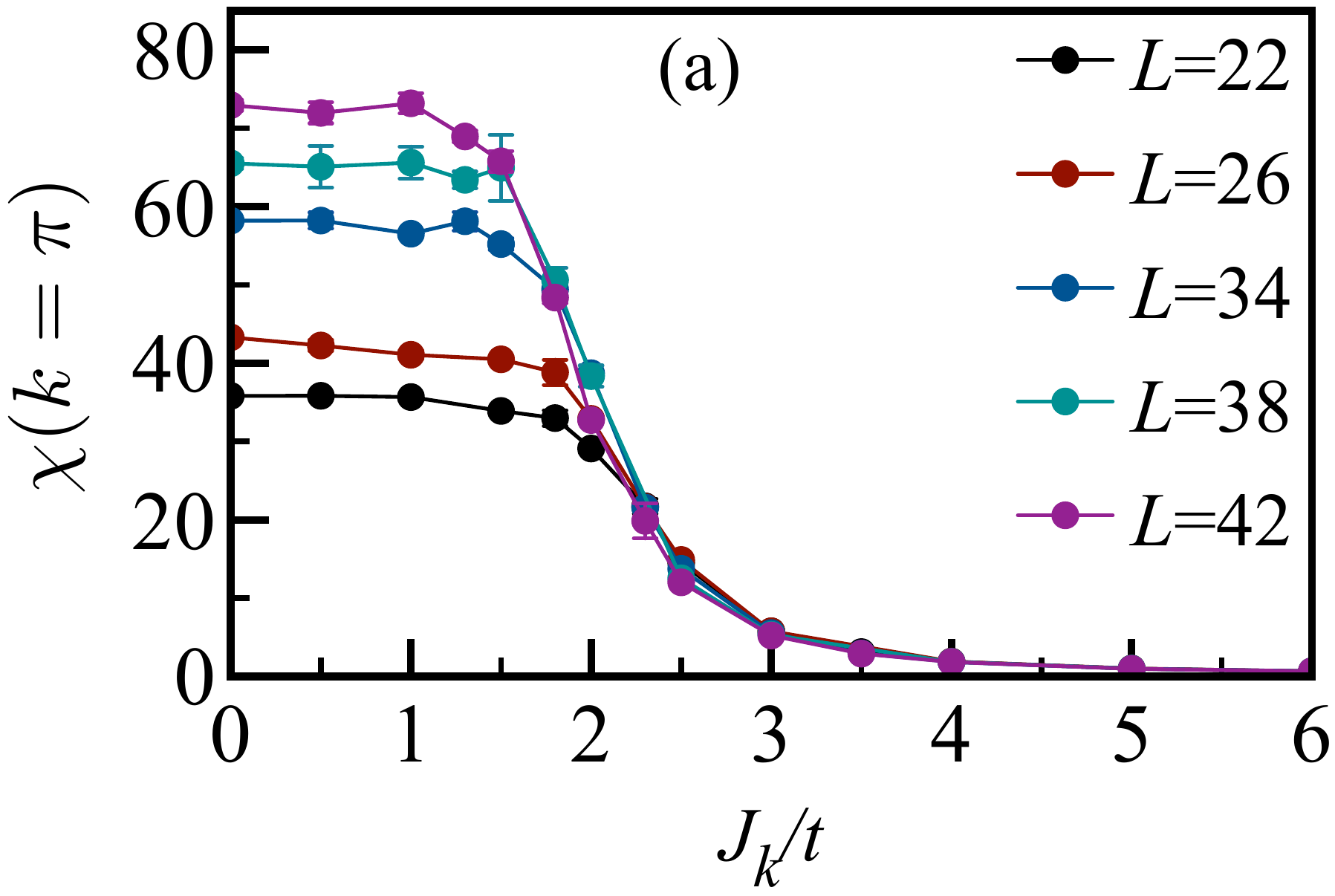}
\includegraphics[width=0.235\textwidth]{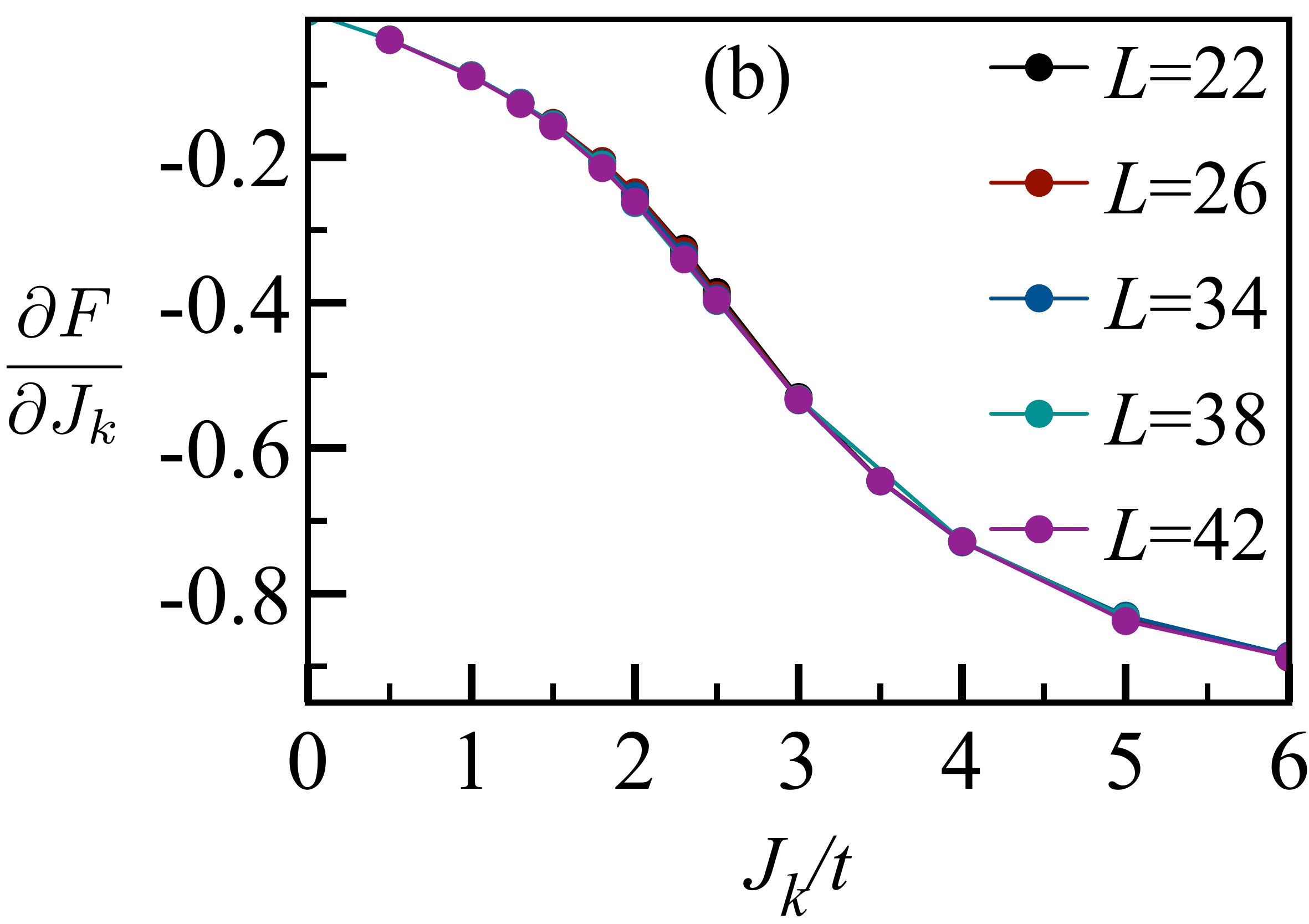}
\caption{Left:  Magnetic susceptibility $\chi(k = \pi)$ as a function of  $J_k/t$  for $J_h/t=1$ and $\beta=L$. Right: Plots $\partial F/\partial J_k$ as a function of $J_k/t$.}
\label{chipi_vs_Jk_piflux}
\end{figure}

To confirm the above, we have computed the spin susceptibility $\chi(k) = \int_{0}^{\beta} d \tau S(k,\tau) $ with $S(k,\tau)$ given as:
\begin{eqnarray}
 S(k,\tau) =   \sum_r e^{-ik \cdot r}  \langle S^z( r,\tau)S^z( r=0,\tau=0) \rangle.
\end{eqnarray}
Lorentz invariance, inherent to spin chains, renders space and time interchangeable such that  the time displaced correlation function scales  as  $1/\sqrt{r^2 + (v_s \tau)^2} $  with $v_s$ the spin velocity.  Setting $\beta = L$,   we  hence expect $\chi (k = \pi )$  to diverge as $L$.  Fig.~\ref{chipi_vs_Jk_piflux} (a)  plots  $\chi (k = \pi )$  at $\beta = L=4n+2$. A similar data at $L=4n+4$ can be found in Fig.~\ref{chipi_vs_Jk_piflux_4n+4} of Ref.~[\onlinecite{suppl}].  For both cases we see  two phases, one in  which  $\chi (k = \pi )$ scales as $L$  and one in which it scales to a $L$-independent constant.  In  Fig.~\ref{chipi_vs_Jk_piflux} (b)  we plot $\frac{1}{L}\frac{\partial F}{\partial J_k}  =\frac{2}{3L}\sum^L_{\ve{l} =1}   \langle  \hat{\ve{c}}^{\dagger}_{\ve{l}}  \ve{\sigma}\hat{\ve{c}}^{}_{\ve{l}} \cdot \hat{\ve{S}}_{\ve{l}} \rangle$ so as to  inquire the nature of the transition. The data favors a smooth curve, and hence a continuous quantum phase transition.

 \begin{figure}[]
\includegraphics[width=0.49\textwidth]{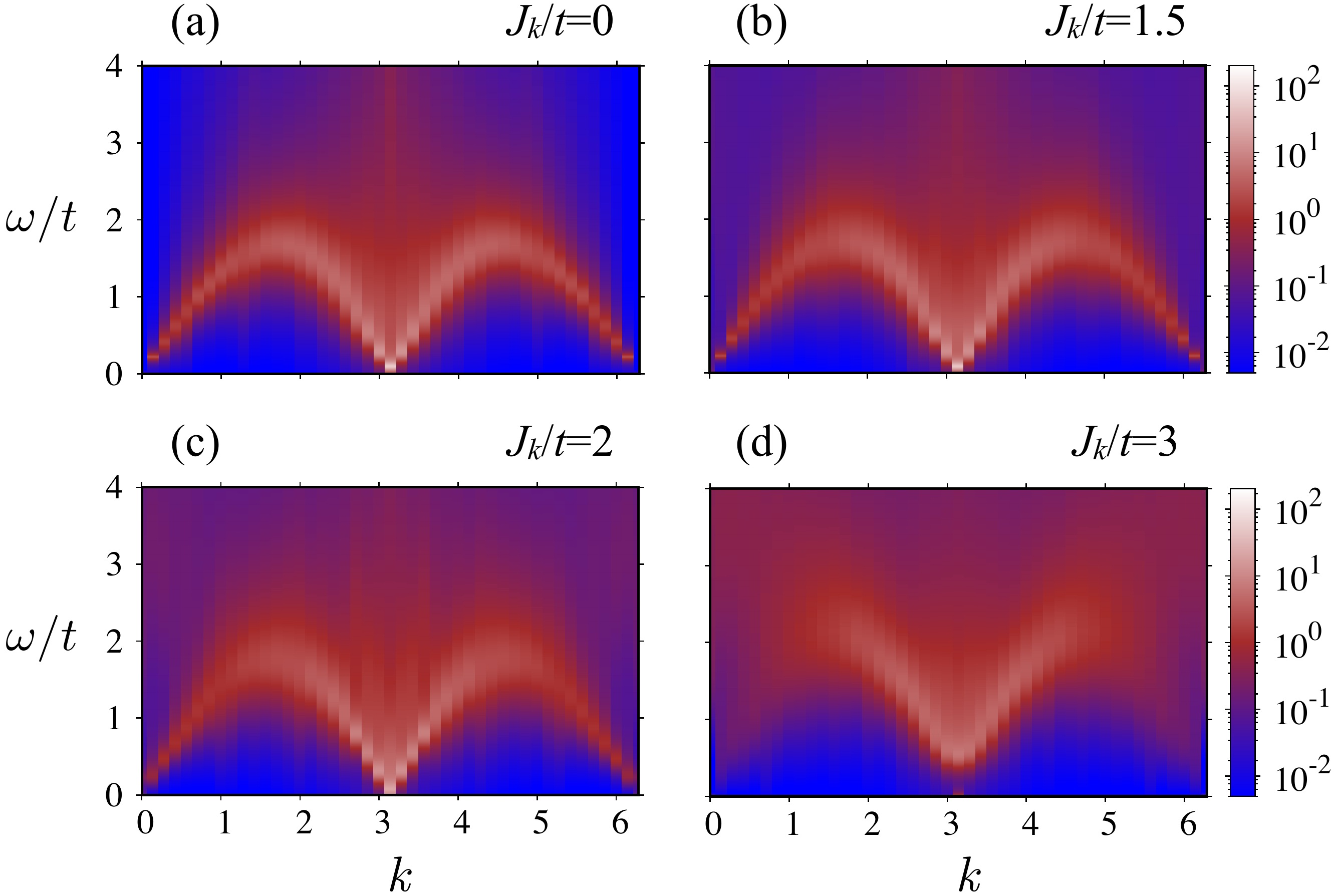}
\caption{Dynamical spin structure factor, $S(k,\omega)$, along spin chain as a function of energy ($\omega/t$) and momentum ($k$) for $L=\beta=44$ at $J_h/t=1$.}
\label{Skomega_piflux_L44_Beta44}
\end{figure}

We now consider the dynamical spin structure factor, that relates to the imaginary-time correlation functions through
$S(k,\tau)  = \frac{1}{\pi} \int d \omega~ \frac{e^{-\tau \omega}}{1-e^{-\beta \omega}}~\chi''(k,\omega)$. To extract $S(k,\omega) = \frac{\chi''(k,\omega)}{1-e^{-\beta \omega} }$, we use the ALF-implementation of the stochastic  analytical continuation algorithm~[\onlinecite{KBeach2004}]. The excitation spectrum of the isolated spin-1/2 Heisenberg chain is  well understood  and  consists of a two-spinon continuum  bounded  by $\frac{\pi}{2} J_h \sin (k) \le \omega (k) \le \pi J_h \sin \left(\frac{k}{2}\right)$. Fig.~\ref{Skomega_piflux_L44_Beta44} plots the dynamical spin spectral function  for different values of $J_k/t$.  Remarkably, the spin dynamics of the Heisenberg chain remains unaffected by conduction electron for $J_k/t\lesssim2$. In  the screened  phase at $J_k/t>2$  spinons bind and low-energy spectral weight is depleted.

 \begin{figure}[]
\includegraphics[width=0.49\textwidth]{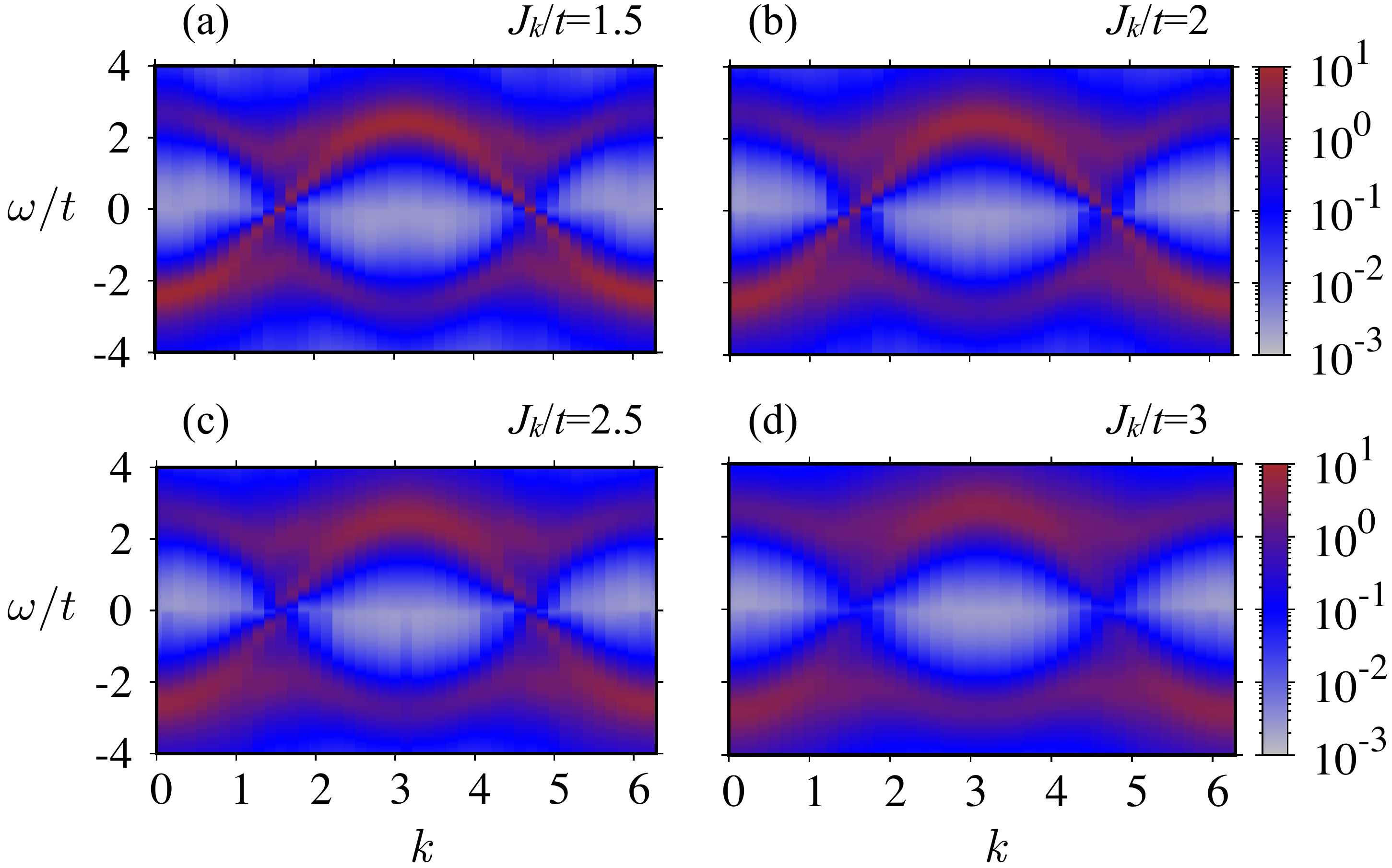}%
\caption{Conduction-electron spectral function, $A_0(k,\omega)$, as a function of energy ($\omega/t$) and momentum ($k$) on  $L=\beta=44$ lattice at $J_h/t=1$.}
\label{Akom_vs_k}
\end{figure}

In Kondo lattices, a Kondo-breakdown transition implies an abrupt change of the Luttinger volume. In our setup such a notion cannot be applied since the localized  spin-1/2 moments are  sub-extensive. Nevertheless, we can consider the spectral function of the conduction electrons that  directly couple to the localized spin-1/2 moments and investigate  how it evolves across the transition. Let $A_n(k,\omega) =-\frac{1}{\pi} \text{Im} G^{\text{ret}} _n(k,\omega)$ with $G^{\text{ret}} _n(k,\omega) = -i \int_{0}^{\infty} d t e^{i \omega  t} \sum_{\sigma} \langle\{\hat{c}_{k,n,\sigma}(0), \hat{c}^{\dagger}_{k,n,\sigma}(t) \}\rangle$.  In the considered Landau gauge, translation symmetry is present along the $x$-direction and $  \hat{c}_{k,n,\sigma} = \frac{1}{\sqrt{L}} \sum_{m=1}^{L}e^{i k m }  \hat{c}_{\ve{i}=(m,n),\sigma}$ is the partial Fourier transform. Fig.~\ref{Akom_vs_k}  plots $ A_0(k,\omega) $ corresponding to the  conduction electrons that  couple  to the Heisenberg chain.  At $J_k=0$  the spectral function shows a dominant $\epsilon(k) = 2t\cos(k a)$ dispersion. In the Kondo-breakdown phase and even at relatively large values of $J_k/t=1.5$ we observe no signs of  hybridization  with the spins. In contrast in the Kondo-screened phase, $J_k/t\gtrsim 2$,  a clear signature of hybridization is apparent.

\begin{figure}[]
\includegraphics[width=0.42\textwidth]{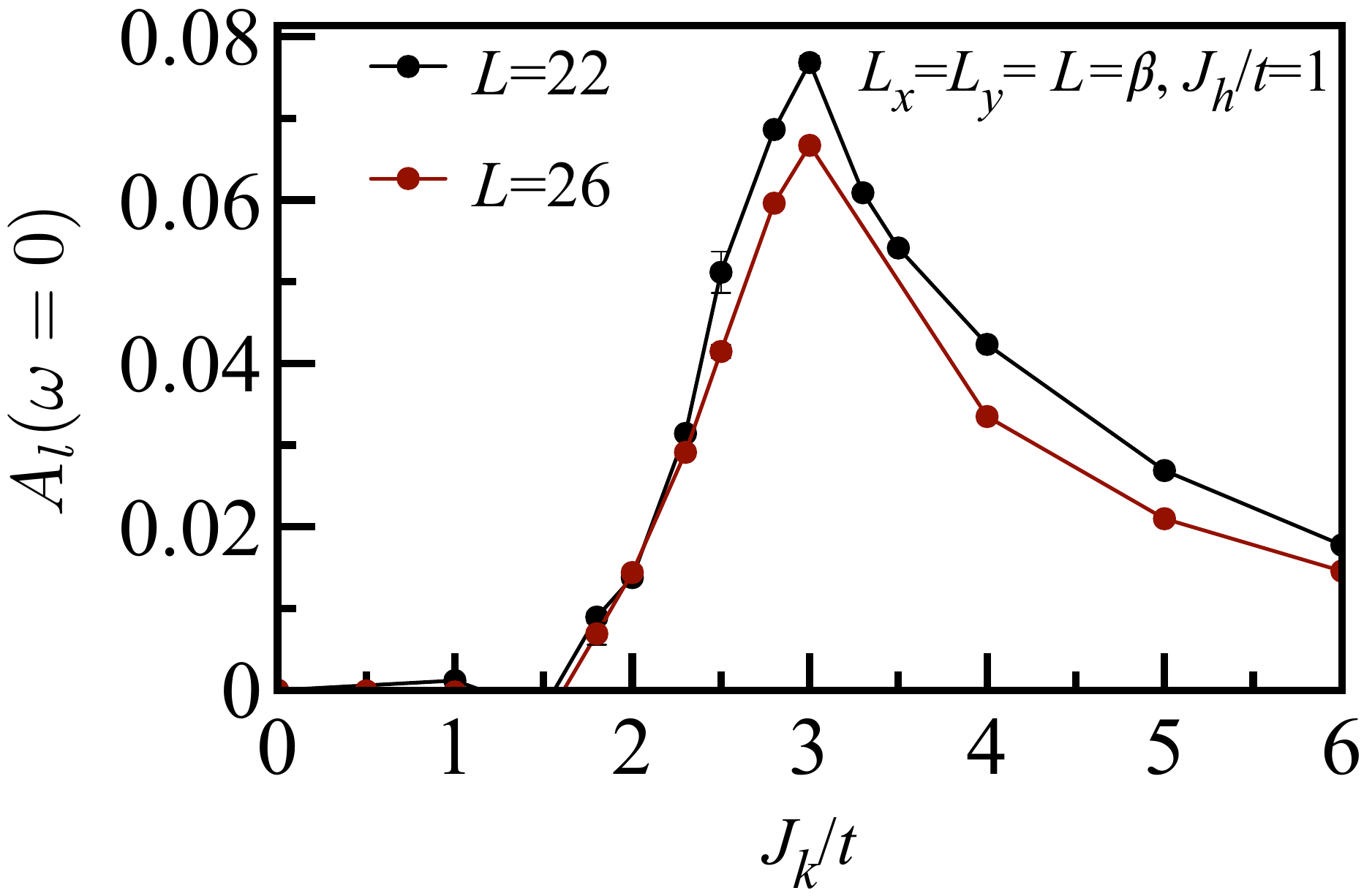}
\caption{ Zero-bias tunneling  through the magnetic adatom.  }
\label{Co_tunneling}
\end{figure}

STM experiments  of magnetic adatoms on metallic  surfaces, separated by an insulating buffer layer shown in Ref.~[\onlinecite{Spinelli2015}, \onlinecite{Toskovic2016}], measure  tunneling  between tip and substrate  occurring through   the localized  orbitals. In our setup  we can access this quantity  by carrying out a Schrieffer-Wolff transformation of  the localized electron creation  operator in the realm of the Anderson model~[\onlinecite{Danu2019}, \onlinecite{Raczkowski18}, \onlinecite{Costi00}].
In particular, $ A_{\ve{l}}(\omega) = - \text{Im}  G_{\ve{l}}^{\text{ret}} (\omega ) $ with  $ G_{\ve{l}}^{\text{ret}}  (\omega) =- i \int_{0}^{\infty} d t  e^{i \omega t }  \sum_{\sigma } \big< \big\{ \tilde{c}_{\ve{l},\sigma}^{}(t), \tilde{c}_{\ve{l},\sigma}^{\dagger} (0) \big\} \big>  $  and  $ \tilde{c}_{\ve{l},\sigma}^{\dagger} =\hat{c}^{\dagger}_{\ve{l}, -\sigma} \hat{S}^{\sigma}_{\ve{l}} + \sigma  \hat{c}_{\ve{l},\sigma}^{\dagger} \hat{S}^{z}_{\ve{l}}$.  Here $\sigma = \pm$   runs over the two spin polarizations and $\hat{S}^{\pm}_{\ve{l}} = \hat{S}^{x}_{\ve{l}}  \pm i \hat{S}^{y}_{\ve{l}} $.  To evaluate the zero-bias  tunneling signal we  estimate $A_{\ve{l}}(\omega=0) \simeq  \frac{1}{\pi}  \beta G_{\ve{l}}(\tau = \beta/2) $. Fig.~\ref{Co_tunneling} plots this quantity.  Remarkably, in the Kondo-breakdown phase, we are not able to distinguish the signal  from zero. This supports the notion that spins and conduction electrons decouple at low energies.  As $J_k \rightarrow  \infty $ the spin binds in a singlet with the conduction electron and the tunneling signal through the adatom drops.
A more detailed numerical analysis~[\onlinecite{Luitz11}, \onlinecite{Karrasch10}] of the STM  signal across the transition is  certainly of great interest.

\textit{Conclusion:} We have shown that a one-dimensional spin  chain  coupled via  a Kondo interaction to 2D Dirac fermions provides a  realization of a continuous Kondo-breakdown  transition. Weak coupling $J_k$ is irrelevant and gapless spinons exist while propagating along the one-dimensional chain. The reason for the absence of Kondo screening in this phase is qualitatively similar to its absence at deconfined quantum critical points in 2D~[\onlinecite{Grover10a}]: in both cases, the anomalous dimension of the spin operator is `large'  due to fractionalization, which makes conduction electrons ineffectual at Kondo screening. Beyond the transition, Kondo screening appears and gapless spinons bind. The Kondo-screened phase is adiabatically connected to the strong-coupling limit, where each spin binds with a conduction electron  into a spin singlet. Larger systems will be needed to determine the critical exponents such as the anomalous dimension of the local moments. In addition, since the number of adatoms in experiments is tunable~[\onlinecite{Toskovic2016,Choi2017,Moro2019}],  it will be very useful to determine how many of them are needed to resolve Kondo breakdown in an interacting spin chain.

The choice of Dirac fermions which only possess Fermi points simplifies the problem and allows  for an RG analysis. This is in contrast to the conventional Hertz-Millis-Moriya approach~[\onlinecite{Hertz76, Millis93, moriya2012spin}] where one integrates out the fermions to obtain an effective non-local action for local moments. Indeed, past work on Fermi surface coupled to a spin-chain employed  Hertz-Millis-Moriya approach, and concluded that the Kondo interaction is relevant (marginal) for an $XXZ$  (Heisenberg) chain, thus destabilizing the Luttinger liquid for infinitesimal Kondo coupling~[\onlinecite{lobos2012magnetic}].  In our problem, the irrelevancy of the Kondo interaction at the decoupled fixed point ($J_k=0$) continues to hold even for a U(1) symmetric $XXZ$ spin-chain and we expect that the qualitative features of our phase diagram will remain unchanged. It will be desirable to study the problem of Fermi surface coupled to an $XXZ$ chain using QMC method, which would also help bridge the gap with experiments in Refs.~[\onlinecite{Spinelli2015,Toskovic2016,Choi2017,Moro2019,Jakob2011,Zhou2010,Serrate2010}]. In addition, other scenarios for Kondo breakdown, such as the one discussed in Ref.~[\onlinecite{komijani2019}], can also be studied using QMC.

In summary, we studied a problem of spin-chain coupled to Dirac fermions and established a Kondo breakdown transition using a combination of techniques. Our results open the window to design and inform new experiments, along the lines of Refs.~[\onlinecite{Toskovic2016,Choi2017,Moro2019}], where adatoms can be suitably arranged on metal/semimetal surfaces.

 \textit{Acknowledgments:} The authors thank M. Aronson, T.-C. Lu,  J. McGreevy for useful conversations, and F. Mila  for insightful collaborations on a related subject.
The authors gratefully acknowledge the Gauss Centre for Supercomputing e.V. (www.gauss-centre.eu) for providing computing time on the GCS Supercomputer SUPERMUC-NG at Leibniz Supercomputing Centre (www.lrz.de).
The research has been supported by the Deutsche Forschungsgemeinschaft through grant number AS 120/14-1 (FFA), the W\"urzburg-Dresden Cluster of Excellence on Complexity and Topology in Quantum Matter - ct.qmat (EXC 2147, project-id 39085490) (FFA and MV), and SFB 1143 (project-id 247310070) (MV). TG is supported by the National Science Foundation under Grant No. DMR-1752417, and as an Alfred P. Sloan Research Fellow. FFA and TG thank the BaCaTeC for partial financial support.

\bibliography{Kondo_ref,fassaad}
\clearpage

\newpage
\widetext
\renewcommand\theequation{S\arabic{equation}}
\renewcommand\thefigure{S\arabic{figure}}
\setcounter{equation}{0}
\setcounter{figure}{0}
\begin{center}
\Large{\bf Supplemental Material for: Kondo breakdown  in a spin 1/2 chain of adatoms  on a Dirac semimetal}\label{meanfield_cal}

{ Bimla Danu,  Matthias Vojta, Fakher F. Assaad and Tarun Grover}

\end{center}
\section{Details of Renormalization Group analysis}\label{RG_cal}
The perturbative RG flow close to the $J_k =0$ fixed point can be obtained via operator product expansion (OPE). We consider Dirac fermions in $d+1$ dimensions coupled to a (1+1)-D spin-1/2 Heisenberg chain. Although the low-energy theory of the Heisenberg chain is not a pure CFT due to marginal operators, we neglect the effect of these operators, and assume that the chain is described by a 1+1-D  SU(2)$_1$ WZW model. Thus, the low-energy imaginary-time action is given by:

\begin{equation}
\mathcal{S} = \mathcal{S}_{\textrm{WZW}} +\int d^dx d\tau\,\,  \overline{\psi} \slashed{p} \psi + J_k \int dx d\tau \,\, \hat{\ve{c}}^{\dagger}(x,\tau) \frac{\ve{\sigma}}{2} \hat{\ve{c}}^{}(x,\tau) \cdot \hat{\ve{S}}(x,\tau)
\end{equation}
where $\mathcal{S}_{\textrm{WZW}} $ is the WZW action for the Heisenberg chain.
The crucial observation is that the tree-level scaling dimension of the operator corresponding to Kondo interaction, $O_{J_k}  = \hat{\ve{c}}^{\dagger}(x,\tau)  \frac{\ve{\sigma}}{2} \hat{\ve{c}}^{}(x,\tau) \cdot \hat{\ve{S}}(x,\tau)$ is $\Delta^{0}_{J_k} = \epsilon$, where $\epsilon = d+1/2 -2 = d-3/2$. This allows for a perturbative access to a UV critical point which becomes unstable towards a stable phase where Kondo interaction is irrelevant (see Fig.~\ref{phases_vs_Jk}). Defining a dimensionless Kondo coupling $j_k=J_k \Lambda^{\epsilon}$, where $\Lambda$ is a UV cut-off scale, the OPE of this operator with itself is given by:
\begin{equation}
:O_{j_k}: :O_{j_k}: = c \identity + :O^2_{j_k}: =c \identity - \frac{1}{2}:O_{j_k}:
\end{equation}
where $:\,: $ denotes normal ordering, $c$ is a constant and $\identity$ denotes the identity operator. The above equation implies that the OPE expansion coefficient $c_{j_k j_k  j_k } = -1/2$ and thus the RG flow equation upto $O(j_k^2)$ is given by:

\begin{equation}
\frac{d j_k}{d \ln \Lambda}=\epsilon j_k - \frac{j^2_k}{2}.
\end{equation}

\section{\bf Auxiliary-field path integral and saddle-point approximation}\label{saddlepoint_QMC}
We consider  SU(N) generalization of the model presented in Eq.({\color{blue}{1}})  of the main text:

\begin{eqnarray}
\Hhat&=&\Hhat_t-\frac{J_k}{4N}\sum_{\ve{l}}  \Big\{ \big(\ve{\hat{c}}^{\dagger}_{\ve{l} }\ve{\hat{d}}^{}_{\ve{l} }+h.c.\big)^2+ \big(i\ve{\hat{c}}^{\dagger}_{\ve{l} }\ve{\hat{d}}^{}_{\ve{l}}+h.c.\big)^2 \Big\}\nonumber\\&&
-\frac{J_h}{4N}\sum_{\ve{l}}  \Big\{ \big(\ve{\hat{d}}^{\dagger}_{\ve{l} }\ve{\hat{d}}^{}_{\ve{l}+\Delta \ve{l}} +h.c.\big)^2+ \big(i \ve{\hat{d}}^{\dagger}_{\ve{l} }\ve{\hat{d}}^{}_{\ve{l}+\Delta \ve{l}} +h.c.\big)^2 \Big\}+\frac{U}{N}\sum_{\ve{l}}   \big(\ve{\hat{d}}^\dagger_{\ve{l}} \ve{\hat{d}}_{\ve{l} }-1\big)^2
\label{ham_sq}
\end{eqnarray}
where  $\ve{\hat{d}}^{\dagger}_{\ve{l}}$ is now an $N$ component fermion (see Ref.~[\onlinecite{Marcin2020}]). Note that since our Monte Carlo simulations are restricted to SU(2)  we can omit the  perfect square terms of the  current operator~[\onlinecite{Assaad99a}, \onlinecite{Capponi2001}].
We will  however pursue  this discussion for a general value of $N$ so as to be able to derive the saddle-point equations in the large-N limit.

Using the Hubbard-Stratonovich (HS) transformation the partition function can be written as:
\begin{eqnarray}
Z~\equiv~\int \mathcal{D}\{\mathcal{V},\mathcal{\chi}, \mathcal{\lambda} \} ~e^{- N \mathcal{S}\{\mathcal{V},\mathcal{\chi},\mathcal{\lambda}\}}
\end{eqnarray}
with the  action
\begin{eqnarray}
\mathcal{S}\{\mathcal{V},\mathcal{\chi},\lambda\}=-\ln \Big[~\mbox{Tr} ~ \mathcal{T} e^{-\int^\beta_0 d\tau~ \Hhat \{\mathcal{V},\mathcal{\chi},\lambda\}}\Big] +\int^\beta_0 d \tau \sum_{\ve{l}} \Big\{\frac{J_k}{4}|\mathcal{V}( \ve{l},\tau)|^2 +\frac{J_h}{4} | \mathcal{\chi}( \ve{l},\tau)|^2+\frac{ U}{4} |\lambda( \ve{l},\tau)|^2\Big\}
 \end{eqnarray}
and  time dependent Hamiltonian
\begin{eqnarray}
 \Hhat \{\mathcal{V},\mathcal{\chi},\lambda\}=\Hhat_t+ \sum_{\ve{l}} \Big\{-\frac{J_k}{2}\big(\mathcal{V}( \ve{l},\tau) \ve{\hat{c}}^{\dagger}_{\ve{l} }\ve{\hat{d}}^{}_{\ve{l} }+h.c\big)-\frac{J_h}{2}\big(\mathcal{\chi}(\ve{l},\tau) \ve{\hat{d}}^{\dagger}_{\ve{l} }\ve{\hat{d}}^{}_{\ve{l}+\Delta \ve{l}}+h.c\big)-i U\lambda(\ve{l},\tau) \big(\ve{\hat{d}}^\dagger_{\ve{l}} \ve{\hat{d}}_{\ve{l} }-1\big) \Big\}.
\end{eqnarray}
 Here, the scalar Lagrange multiplier $\mathcal{\lambda}( \ve{l},\tau)$ enforces the constraint and  $\mathcal{V( \ve{l},\tau)}$  and $\mathcal{\chi( \ve{l},\tau)}$  are complex  bond  fields. In the  QMC  simulations we sum over all field configurations to obtain an exact result. At the particle-hole symmetric point, and as argued in the main text, the imaginary part of the action takes the value  $ n \pi$ with $n$ an integer. Hence for even  values of $N$  no negative sign problem occurs.

 In the large-N limit,  we expect the saddle-point approximation to become exact:
\begin{eqnarray}
\frac{d\mathcal{S}\{\mathcal{V},\mathcal{\chi},\mathcal{\lambda}\}}{d \mathcal{V}( \ve{l},\tau)},\frac{d\mathcal{S}\{\mathcal{V^*},\mathcal{\chi^*},\lambda\}}{d \mathcal{V}^*( \ve{l},\tau)}=0, \quad \mbox{and}, \quad\frac{d \mathcal{S}\{\mathcal{V},\mathcal{\chi},\mathcal{\lambda}\}}{d \mathcal{\chi}( \ve{l},\tau)},\frac{d\mathcal{S}\{\mathcal{V^*},\mathcal{\chi^*},\mathcal{\lambda}\}}{d \mathcal{\chi}^*( \ve{l},\tau)}=0,  \quad \mbox{and},\quad\frac{d\mathcal{S}\{\mathcal{V},\mathcal{\chi},\mathcal{\lambda}\}}{d \mathcal{\lambda}( \ve{l},\tau)}=0.
\end{eqnarray}

In the mean-field approximation carried  out in the next sections,   we restrict the  search  for saddle points to  space and time independent fields:
 $ \mathcal{V}( \ve{l},\tau)=\mathcal{V}^*( \ve{l},\tau)=V(\in R)$ and $\mathcal{\chi}( \ve{l},\tau)=\mathcal{\chi}^*( \ve{l},\tau)=\chi(\in R)$  and impose the constraint only on average.
\section{\bf Large-N mean field calculation for a chain of magnetic adatoms} \label{meanfield_cal}
We consider an infinite  Heisenberg chain of adatoms with periodic  boundary conditions. The unit cell, denoted as $l$, contains $n \in \left[ 1 \cdots N_c \right] $ conduction electrons  $ \hat{c}_{l,n,\sigma} $ and a single spin degree of freedom. In this case the mean-field Hamiltonian can be written as,
\begin{eqnarray}
\hat{H}^{}_{mf} =& & \sum_{k,\sigma,n,n'} \hat{c}^{\dagger}_{k,n,\sigma} T(k)_{n,n'} \hat{c}^{}_{k,n^\prime,\sigma}-\frac{J_h\chi}{2}\sum_{k,\sigma} {\epsilon}^d_{k}\hat{d}^{\dagger}_{k,\sigma}  \hat{d}^{}_{k,\sigma}+\lambda\sum_{k,\sigma}\hat{d}^{\dagger}_{k,\sigma}  \hat{d}^{}_{{k},\sigma}-\frac{J_kV}{2} \sum_{k,\sigma}(\hat{c}^{\dagger}_{k,1,\sigma}  \hat{d}_{k,\sigma} + h.c)+e_0N_{u}
\end{eqnarray}
where, ${\epsilon}^d_{k}=2\cos(ka)$, $e_0=(\frac{J_kV^2} {2}+\frac{J_h\chi^2} {2}-\lambda+\mu)$  and   $N_u$ is the number of unit cells.   Hereafter, we will set $a=1$.  The mean-field order parameters are defined as $ V =\sum_{\sigma} \langle \hat{c}^{\dagger}_{l,1,\sigma}  \hat{d}^{}_{l,\sigma}  \rangle=\sum_{\sigma} \langle \hat{d}^{\dagger}_{l,\sigma}  \hat{c}^{}_{l,1,\sigma}  \rangle$,  $\chi = \sum_\sigma \langle  \hat{d}^{\dagger}_{l+1,\sigma} \hat{d}^{}_{l,\sigma} \rangle = \sum_\sigma \langle  \hat{d}^{\dagger}_{l,\sigma} \hat{d}^{}_{l+1,\sigma} \rangle$, and, the Lagrange multiplier $\lambda$   will enforce the constraint  $\sum_\sigma \langle  \hat{d}^{\dagger}_{l,\sigma} \hat{d}^{}_{l,\sigma} \rangle = 1$ on average.
The  mean field Hamiltonian  is then given by:
\begin{eqnarray}
\hat{H}_{mf}=\sum_{k,\sigma} \hat{\ve{\phi}}^\dagger_{k,\sigma} M(k)  \hat{\ve{ \phi}}_{k,\sigma }+N_ue_0
\end{eqnarray}
with  $\ve{\hat{\phi}}^\dagger_{k,\sigma} =\left(\hat{c}^{\dagger}_{k,1,\sigma}, \cdots,  \hat{c}^{\dagger}_{k,N_c,\sigma}, \hat{d}^{\dagger}_{k,\sigma}  \right) $  and

\begin{eqnarray*}
& & M(k)=  \\
& & \left( \begin{array}{cccccccccccccccc}
-2t \cos k - \mu&  ~~   -t-2t^\prime \cos  k &       0    &  0&0&  \cdots& \cdots&~  -t-2t^\prime \cos  k&-\frac{J_k V}{2}\\
-t -2t^\prime \cos  k&  ~~  2t \cos k - \mu&     ~~     -t -2t^\prime \cos  k &  0& 0&  \cdots&   \cdots &0&0\\
0&     ~~ -t -2t^\prime \cos  k  &  ~~    -2t \cos k - \mu& ~~ -t -2t^\prime \cos  k & 0 &   \cdots&  \cdots&0&0\\
\cdots&       \cdots&         \cdots&  \cdots& \cdots&\cdots&    \cdots& \cdots&   \cdots\\
\cdots&       \cdots&         \cdots&  \cdots& \cdots&\cdots&    \cdots& \cdots&   \cdots\\
-t -2t^\prime \cos  k  ~~&      0&      0& 0&0&  \cdots& \cdots&   2t \cos k - \mu &  0\\\
 -\frac{J_k V}{2}&0&0&0&0& \cdots& \cdots&0&- J_h \chi\cos k + \lambda
\end{array}
\right).
\end{eqnarray*}
In the above  we have added a next-nearest hopping $t'$ term that lifts the particle-hole (ph) symmetry of the model. We will nevertheless  consider  the half-filled case. Diagonalising:
$  U^{\dagger}(k) M (k)  U(k)  =  \text{Diag} \left(  E_{k,1},   \cdots,  E_{k,N_c + 1 } \right)$ gives:
\begin{eqnarray}
\hat{H}_{mf}=N_{u}e_0+\sum_{k,\sigma} \sum^{N_{c+1}}_{n=1}E_{k,n} {\hat{\psi}}^\dagger_{k,n,\sigma} {\hat{\psi}}_{k,n,\sigma}
\end{eqnarray}
with $\ve{\hat{\psi}}^\dagger_{k,\sigma} = \ve{\hat{\phi}}^\dagger_{k,\sigma} U(k) $.

 The ground state energy per unit cell can be computed as
\begin{eqnarray}
e_g= e_0+\frac{2}{N_{u}}\sum_{k} \sum^{N_{c+1}}_{n=1,  E_{k,n}  < 0 }E_{k,n}.
\label{Egkn_vs_k}
\end{eqnarray}
The mean-field parameters $V$  and $\chi$ are obtained by minimizing the  ground state energy ($e_g$) using the simplex method (see Ref.~[\onlinecite{Nelder1965}])  in the  constrained space  $\sum_\sigma \langle  \hat{d}^{\dagger}_{l,\sigma} \hat{d}^{}_{l,\sigma} \rangle = 1$ and  total  number of particles  $L^2 + L$. These constraints are imposed by adjusting  $\mu$ and $\lambda$.

\subsection{\bf The particle-hole symmetric case}
\begin{figure}[htbp]
\includegraphics[width=0.78\textwidth]{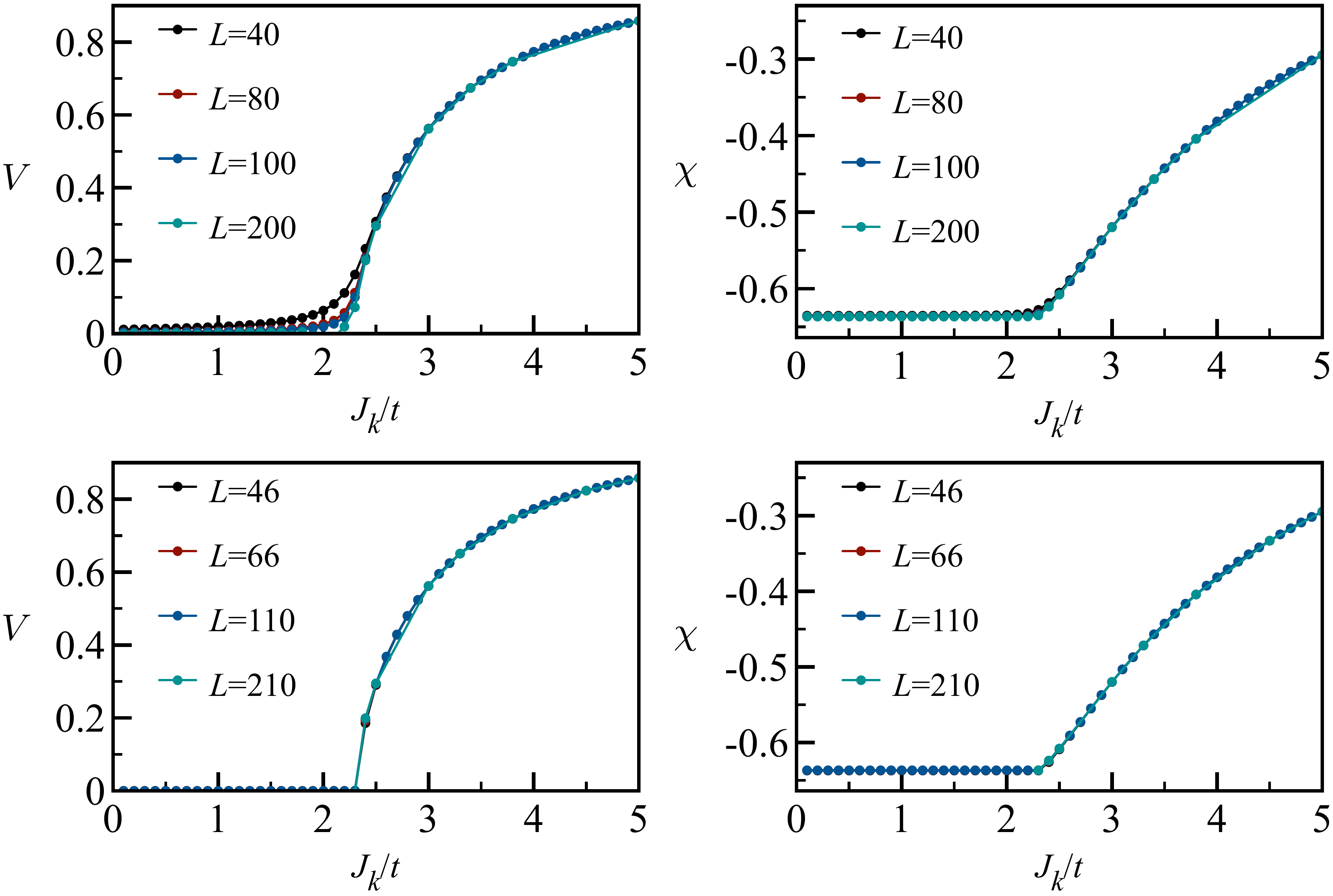}
\caption{Top panel: Mean-field parameters $V$ and $\chi$ as a function of  $J_k/t$ at $J_h/t=1$ on $4n+4$ lengths chain. Bottom panel: Mean-field parameter $V$ and $\chi$ as a function of $J_k/t$ at $J_h/t=1$ on $4n+2$ lengths chain. In both case the we have use periodic boundary condition along the spin chain direction as well in $y$-direction of the substrate.}

~

\label{V_chi_vs_Jk_piflux_per}
\includegraphics[width=0.78\textwidth]{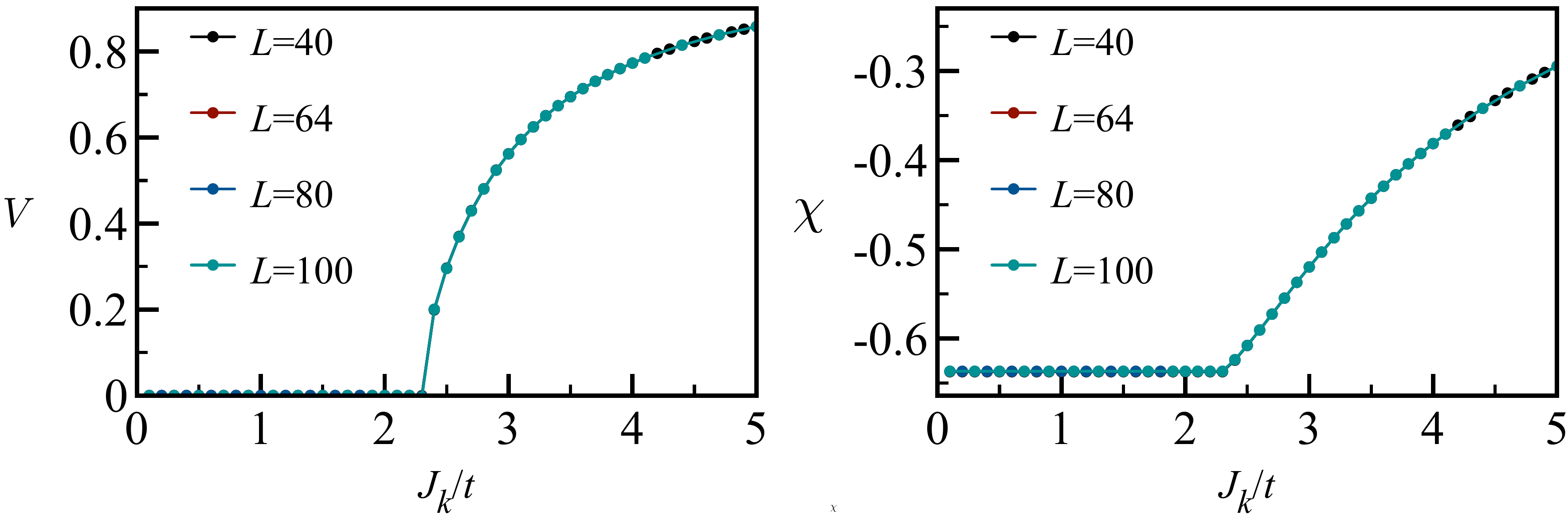}
\caption{Mean-field parameters $V$ and $\chi$ as a function of  $J_k/t$ at $J_h/t=1$ for $L=4n+4$  with anti-periodic boundary condition. }
\label{V_chi_vs_Jk_piflux_antiper}
\end{figure}

At $t'=0$  both the chemical potential and   Lagrange  multiplier  are  pinned to zero due to particle-hole symmetry.
The mean field   values  for $ V$ and $\chi$ as a function of $J_k/t$  at fix $J_h/t=1$ are shown in  Fig.~\ref{V_chi_vs_Jk_piflux_per} for $L=4n+4$ (see top panel)  and for $ L=4n+2$ (see bottom panel).  
For one  dimensional systems, it is advantageous  to adapt the boundary condition to the particle number  so as to optimize  extrapolation to the thermodynamic limit.  In particular,  anti-periodic (periodic) boundaries  for systems with 4n+4 (4n + 2) guarantees that the non-interaction system has a  unique ground state.  As apparent  in Figs.~\ref{V_chi_vs_Jk_piflux_per}  (bottom) and \ref{V_chi_vs_Jk_piflux_antiper}, when these conditions are met,  a quick scaling to the thermodynamic limit is obtained.  If not  (see Fig.~\ref{V_chi_vs_Jk_piflux_per}  (top))  extrapolation to the thermodynamic limit is hard, but ultimately, the same results are obtained.

The mean-field results clearly show two phases upon  tuning $J_k/t$   at fixed $J_h$.  The decoupled  phase is characterized by $V=0$, $\chi \ne 0 $, and the  Kondo-screened state  by $V\ne0$, $\chi \ne 0$.   The mean-field phase diagram  in  the $J_h/t$  versus $J_k/t$ plane  is shown in Fig.~\ref{meanfld_phase_Jh_Jk} of main text.   The transition is continuous  and,  the  mean-field Kondo breakdown critical point  takes place at  $J^c_k/t\approx2.3$ for  $J_h/t=1$.

\begin{figure}[h]
\includegraphics[width=0.86\textwidth]{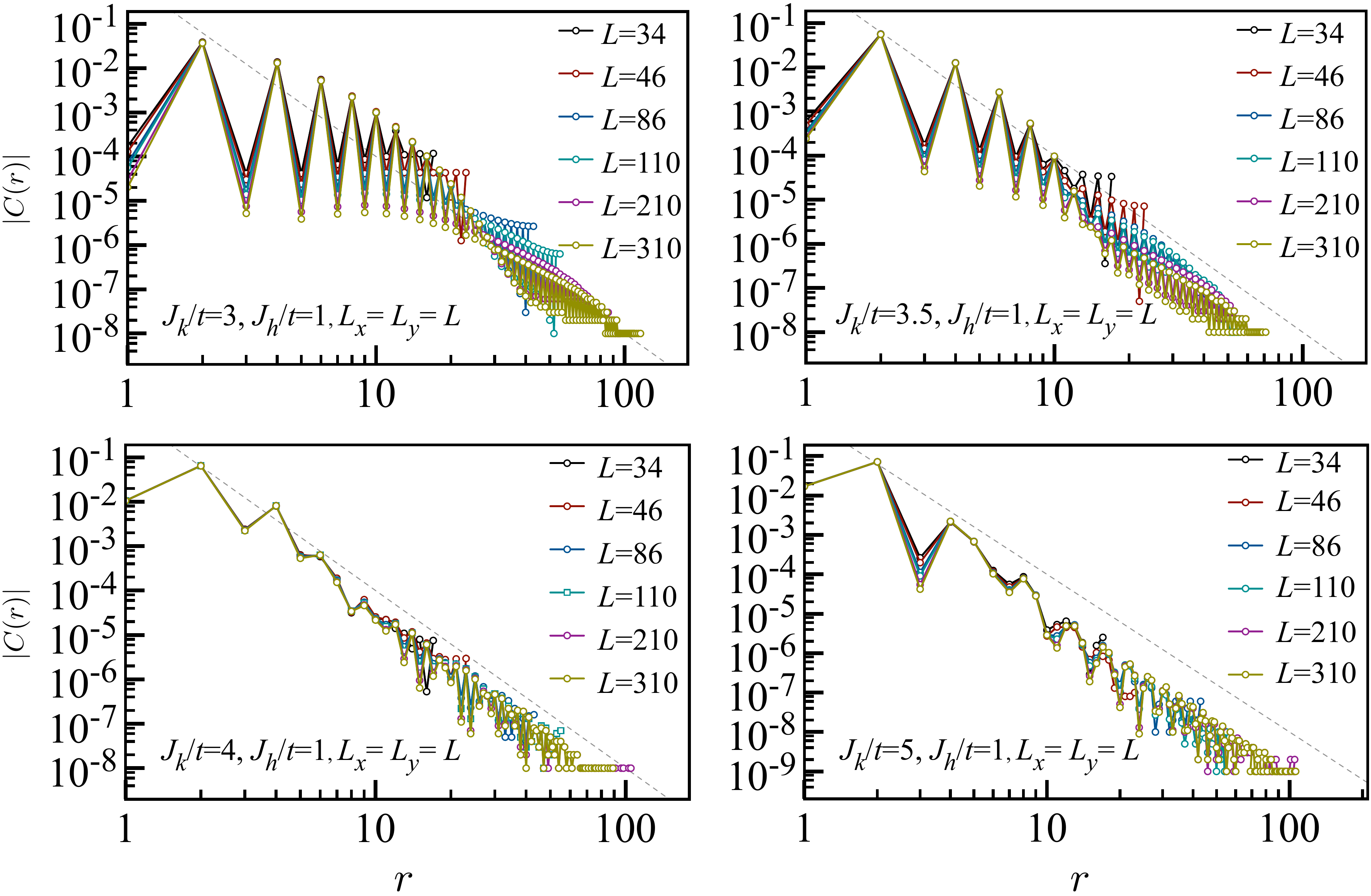}
\caption{ Log-log scale plots of equal time spin-spin correlation function $C(r)$ as a function of distance $r$  along the spin chain for the given  $J_k/t$ values computed in large-N mean-field calculation. The grey dashed line  corresponds to the $1/r^4$  form. }
\label{meanfield_correlation}
\end{figure}

 In the decoupled phase,  the spin-spin correlations,  $C(r)=\frac{1}{L} \sum_q e^{-iq.r} \langle  S^z(q)  S^z(-q)\rangle$  decay as $1/r^2$  reflecting the scaling dimension  $d/2$  with $d=1$   for the mean-field description of spinons. In the Kondo-screened phase we expect the spin-spin correlations along the chain to acquire the  scaling behavior  of the Dirac  substrate, $1/r^4$. This expectation, that stems from the fact that the spin system is sub-extensive and hence cannot  alter the properties of the substrate,  is positively checked in Fig.~\ref{meanfield_correlation}.

\begin{figure}[htbp]
\includegraphics[width=0.97\textwidth]{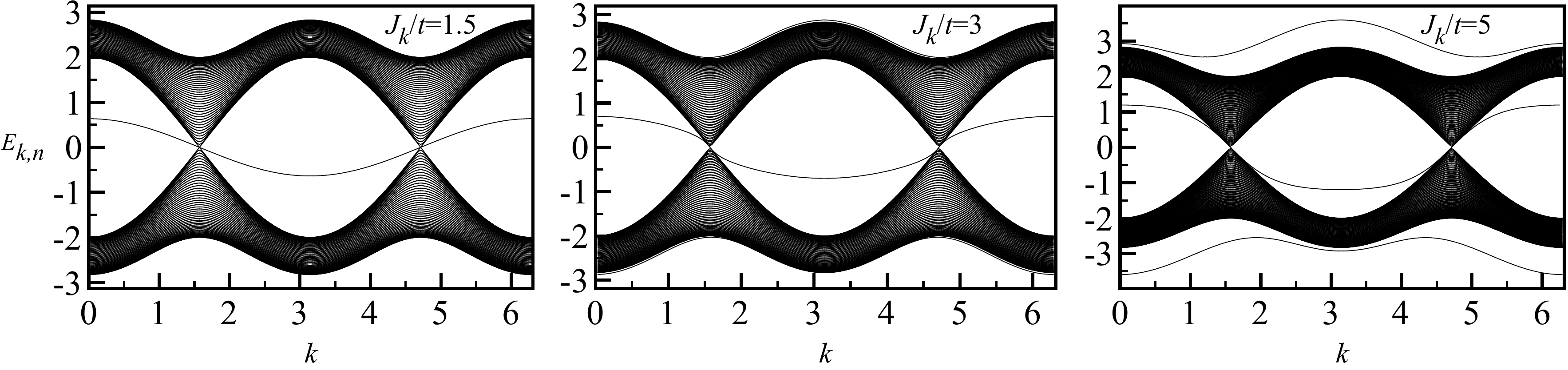}
\caption{Band energies, $E_{k,n}$, as a function of momentum ($k$) for $J_h/t=1$ and $L_x=L_y=L=200$. Particularly, in the decoupled  (left), Kondo-screened (center)  and strong-coupling (right) phases.   Here,  particle-hole symmetry results in:
$E_{k+\pi,n} = - E_{k,n'} $   for  a pair of bands $n,n'$.}
\label{Ek_vs_k_L200_piflux}
\end{figure}

 Finally in Fig.~\ref{Ek_vs_k_L200_piflux} we plot   $E_{k,n}$ in  the  decoupled and Kondo-screened phases. In the decoupled phase, $J_k/t = 1.5$, we observe the spinon-band and the 2D Dirac electrons. With our gauge choice and  periodic boundary conditions, the Dirac cones are located at $\ve{k} =  \pm \left(\frac{\pi}{2}, \frac{\pi}{2} \right) $.  Furthermore, the Fermi velocity is set  by $v_F = 2t$ and the spinon velocity by $J_h \chi$.  In the  Kondo-screened phase at $J_k/t = 3$  hybridization between spin  and  Dirac electrons is  apparent.  In the limit  of large $J_k$   we observe  bonding and  anti-bonding bands  of  the spinon and conduction electrons  $\hat{c}_{k,1,\sigma}$  that split off  at  low and high energy.   In this limit, the Dirac  electrons are subject to open boundary conditions in the $y$-direction. For  the given cut,  no edge states  are  expected.

\subsection{\bf The particle-hole asymmetric case}\label{meanfield_cal_aph}
In free standing graphene, particle-hole symmetry is an emergent symmetry.  The question we will  address here is if our results  depend on  the Hamiltonian being particle-hole symmetric.  The breaking of this symmetry  results in a negative sign problem  in the QMC approach,  such that we will answer this question in the realm of  the large-N mean-field theory.  Including a $t'$  matrix  element in our  calculations breaks the particle-hole symmetry.  Since the  point group symmetry  remains unchanged, the  Dirac  cones do not meander  and remain pinned at   $\ve{K} = \pm(\pi/2, \pi/2) $  for our Landau gauge choice.

\begin{figure}[htbp]
\includegraphics[width=0.97\textwidth]{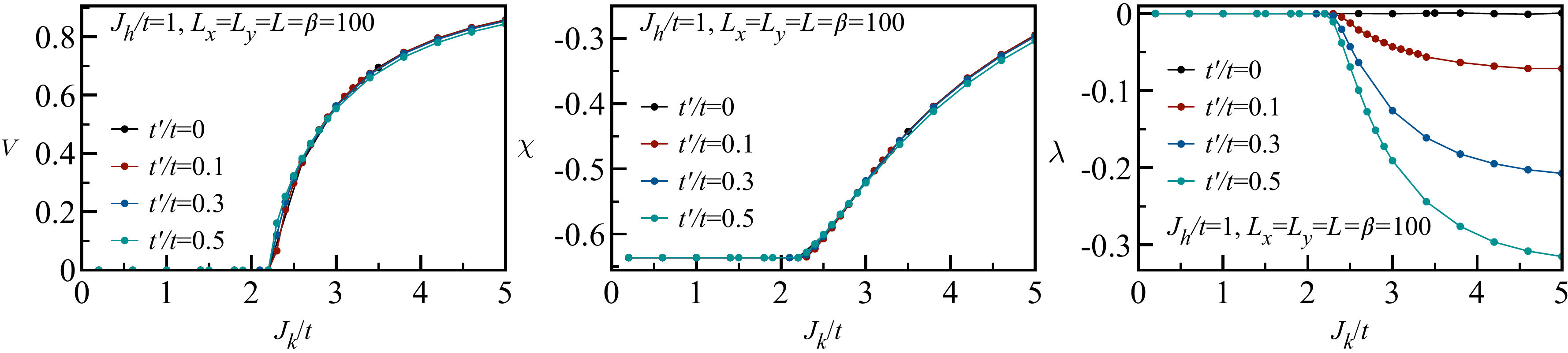}
\caption{Mean-field parameters; $V$ (left), $\chi$ (center) and $\lambda$ (right), as a function of  $J_k/t$ for the given parameters values.}
\label{V_chi_lambda_vs_Jk_piflux_aph}

~

\includegraphics[width=0.97\textwidth]{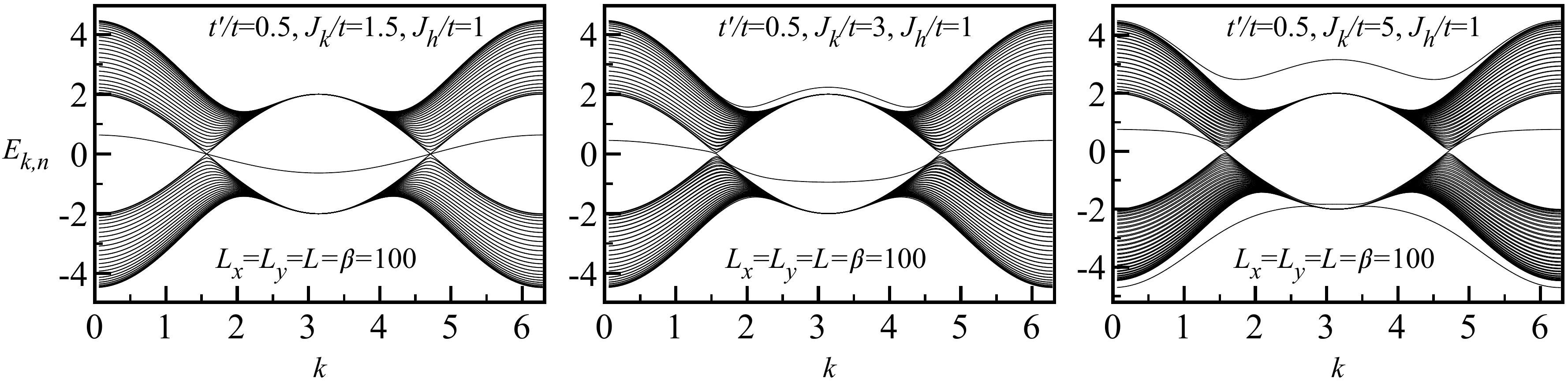}
\caption{Band energies, $E_{k,n}$, as a function of momentum ($k$) for the given parameters values. Particularly, in the decoupled  (left), Kondo-screened (center) and strong-coupling (right) phases.}
\label{Ek_vs_k_L200_piflux_aph}
\end{figure}

The mean-field order parameters  as well as the Lagrange multiplier are plotted in Fig.~\ref{V_chi_lambda_vs_Jk_piflux_aph}  for various values of $t'$.  Remarkably, the  results remain next to unchained. To understand whay, it is instructive to analyze low-energy effective model  for the Dirac electrons at finite $t'$.  The corresponding tight-binding Hamiltonian with a nearest ($t$) and  the next-nearest neighbor ($t^\prime$)  hopping on the square  lattice  on a torus,    Landau gauge $\ve{A}=B(-y,0,0)$  with  $B a^2 /\phi_0 =  1/2 $,  can be written as,
\begin{eqnarray}
\hat{H}^{t t^\prime}=\sum_\k \big( \begin{array}{cccccccc}
 \hat{a}^\dagger_\k, & \hat{b}^\dagger_\k \\
  \end{array}
\big)  \left(\begin{array}{cccccccc}
-2t \cos k_x &   \underbrace{-t( 1+ e^{- i 2 k_y})  -2t^\prime  \cos k_x ( 1+ e^{-2 i k_y})}_{Z_\k} \nonumber \\
 \underbrace{-t( 1+ e^{i 2 k_y}) -2t^\prime  \cos k_x ( 1+ e^{2 i k_y})}_{Z^*_\k}     & 2t \cos k_x
\end{array}
\right)
\left( \begin{array}{cccccccc}
\hat{a}_\k\\
\hat{b}_\k \\
 \end{array}
\right).
\\
\label{piflux_ham}
\end{eqnarray}
In the above we have  set $a=1$.  The magnetic unit cell    contains two orbitals with associated creation operators, $\hat{a}^{\dagger}_{\k}$  and $\hat{b}^{\dagger}_{\k}$,  and the lattice  vectors are given by $ \ve{a}_1  = (1,0)  $,  $ \ve{a}_2  = (0,2) $.
The dispersion then reads:
\begin{eqnarray}
\epsilon_{\pm}(\k)= \pm \sqrt{ 4t^2\cos^2 k_x+ |Z(\k)|^2}.
\label{epsilon_piflux}
\end{eqnarray}
By symmetry one  will check that the Dirac points  remain  pinned at  $\K=\pm \frac{\pi}{2}(1,1)$.
Expansion  around the Dirac points gives:
\begin{eqnarray}
H^{tt^\prime}(\pm \K+\q)\approx2 t \big( \pm \hat{\sigma}_z q_x -\hat{\sigma}_y q_y\big)\mp 4 t^\prime q_x q_y \hat{\sigma}_y.
\label{epsilnK+q}
\end{eqnarray}
As apparent in the low-energy limit,  $t^\prime$,   can  be neglected since it comes  in second order in $\ve{q}$.  This explains  the fact that our results, at least at weak coupling remain unchanged.

The  band energies, $E_{k,n}$, as a function of momentum $k$ are shown in Fig.~\ref{Ek_vs_k_L200_piflux_aph}.   Here, the  particle hole asymmetry is apparent since
the relation $ E_{k+\pi,n} = -  E_{k,n'} $  does not hold for a pair of bands $n,n'$.

~

\newpage

~

\section{Details of QMC results}\label{QMC_data_details}
Here we provide additional QMC  results.
 \begin{figure}[htbp]
\includegraphics[width=0.92\textwidth]{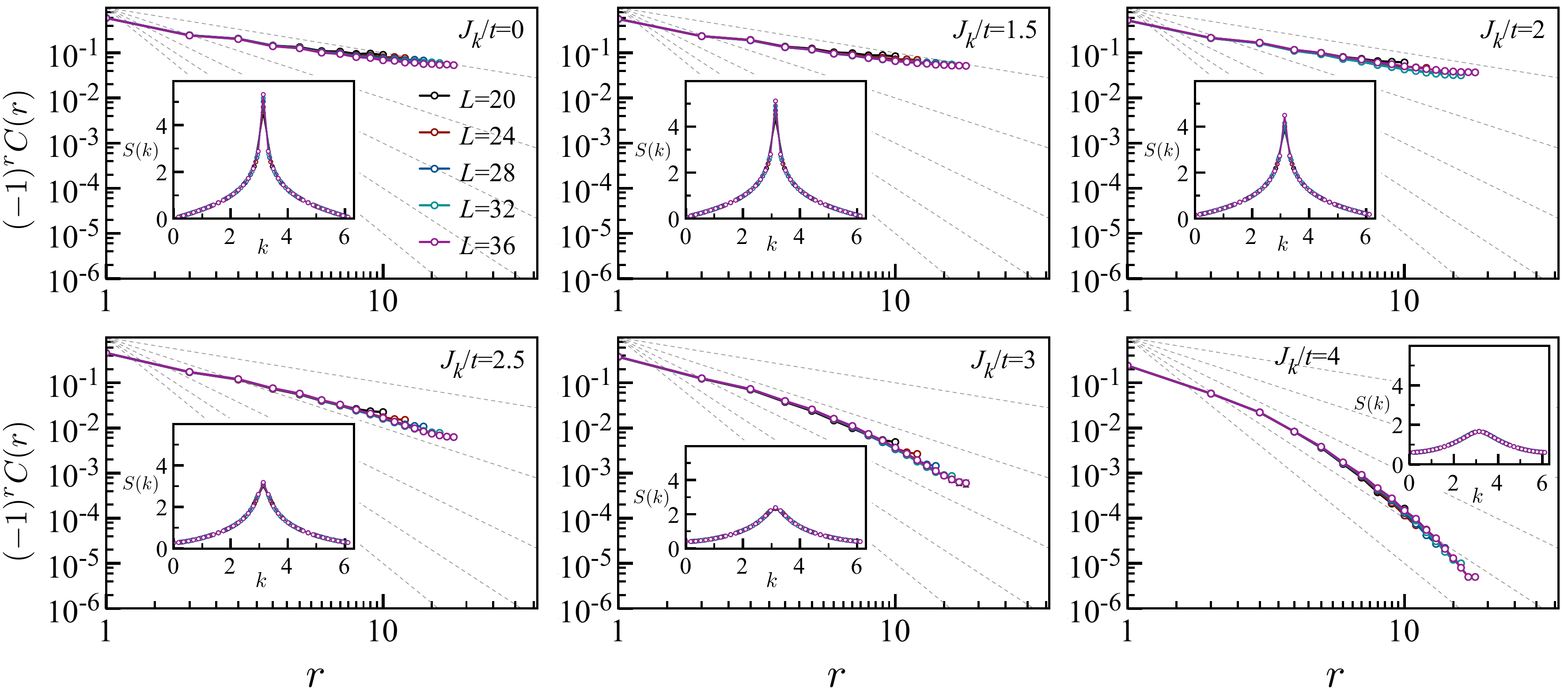}%

\caption{Equal time spin correlation function $C(r)$  as a function of distance $r$ in log-log scale for given  values of $J_k/t$  on $4n+4$ lengths chain  for $J_h//t=1$ and $L_x=L_y=L=\beta$. The corresponding static spin structure factors $S(k)$ as a function of momentum ($k$) are shown in subfigures. The grey dashed line denotes the $1/r^n$ with powers $n$ starting from $n=1$ (first line)  to $n=5$ (last line).}
\label{LogSr_vs_logr_4n+4_dtaup05}
\end{figure}
 \begin{figure}[htbp]
\includegraphics[width=0.5\textwidth]{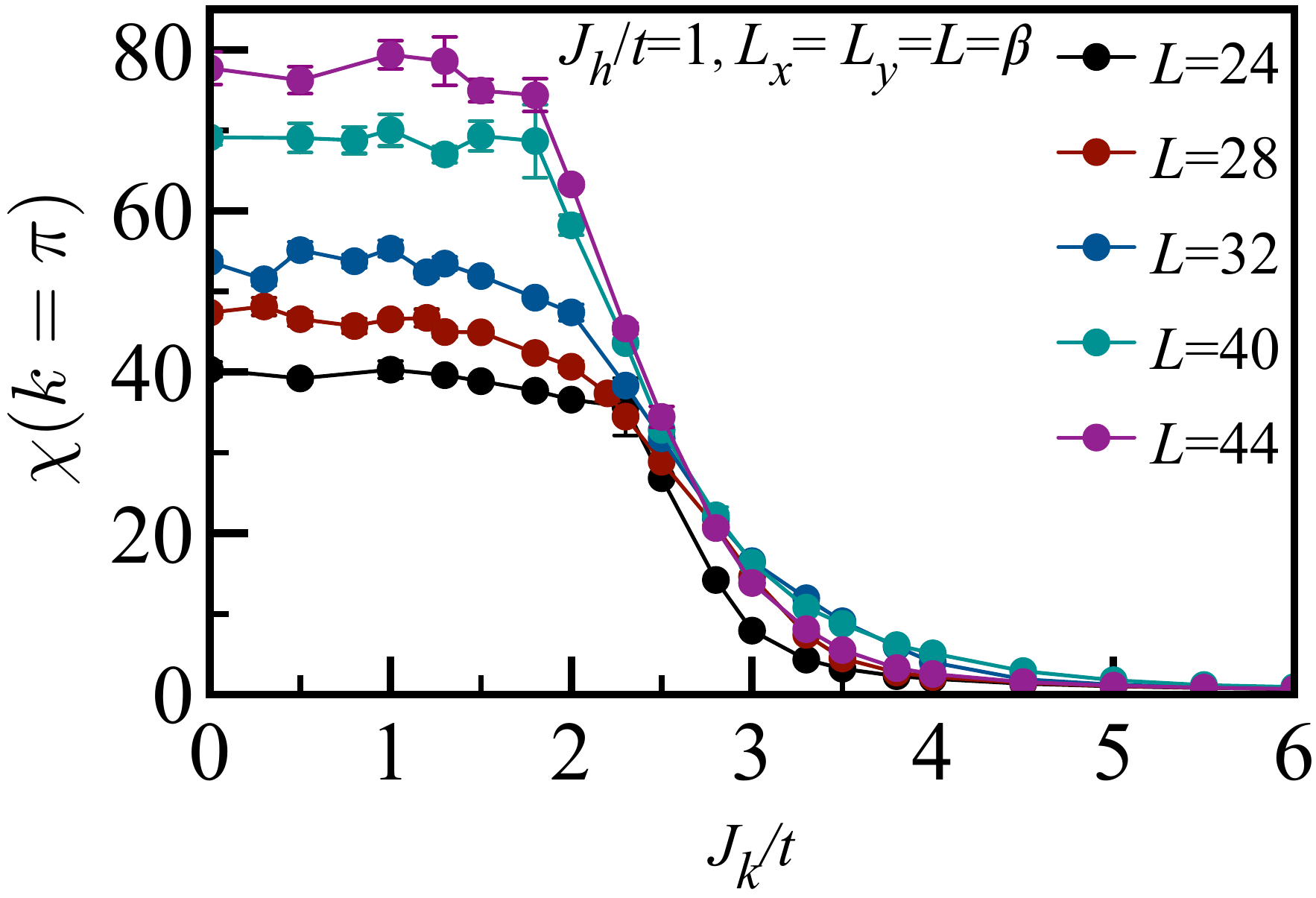}
\caption{Magnetic susceptibility, $\chi(k = \pi)$, as a function of  $J_k/t$ for different system size at $\beta=L$ on $4n+4$ lengths chain.}
\label{chipi_vs_Jk_piflux_4n+4}
\end{figure}

 \begin{figure}[htbp]
\includegraphics[width=0.95\textwidth]{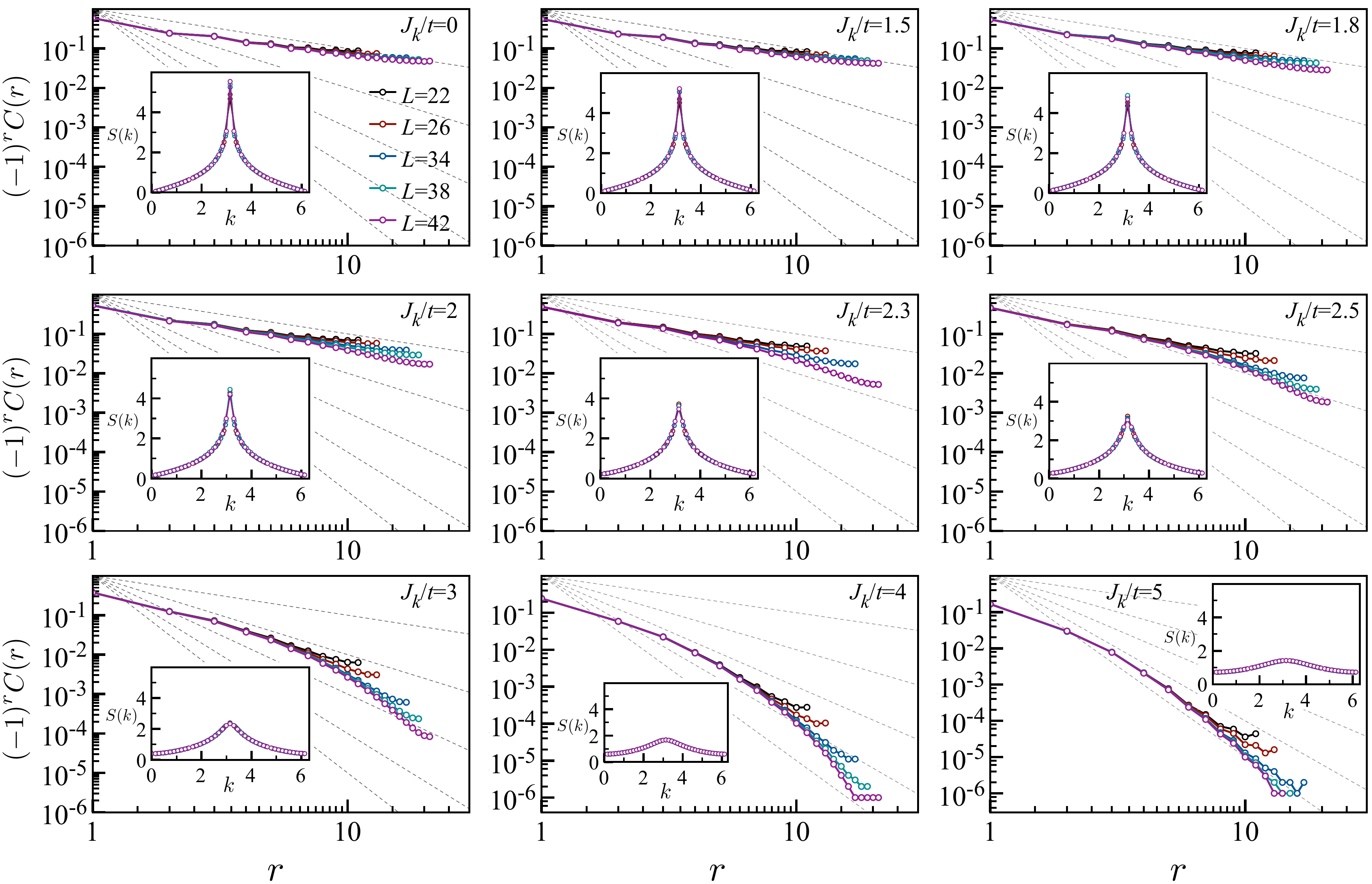}%
\caption{More details of equal time spin correlation shown in Fig.~\ref{LogSr_vs_logr_4n+2} of the main paper. Equal time spin correlation function, $C(r)$, as a function of distance $r$ in log-log scale for various values of $J_k/t$ on $4n+2$ lengths chain  for  $J_h//t=1$ and $L_x=L_y=L=\beta$. The corresponding static spin structure factors $S(k)$ as a function of momentum ($k$) are shown in subfigures. The grey dashed line denotes the $1/r^n$ with powers $n$ starting from $n=1$ (first line)  to $n=5$ (last line).}
\label{LogSr_vs_logr_4n+2_all}
\end{figure}

 \begin{figure}[htbp]
\includegraphics[width=0.96\textwidth]{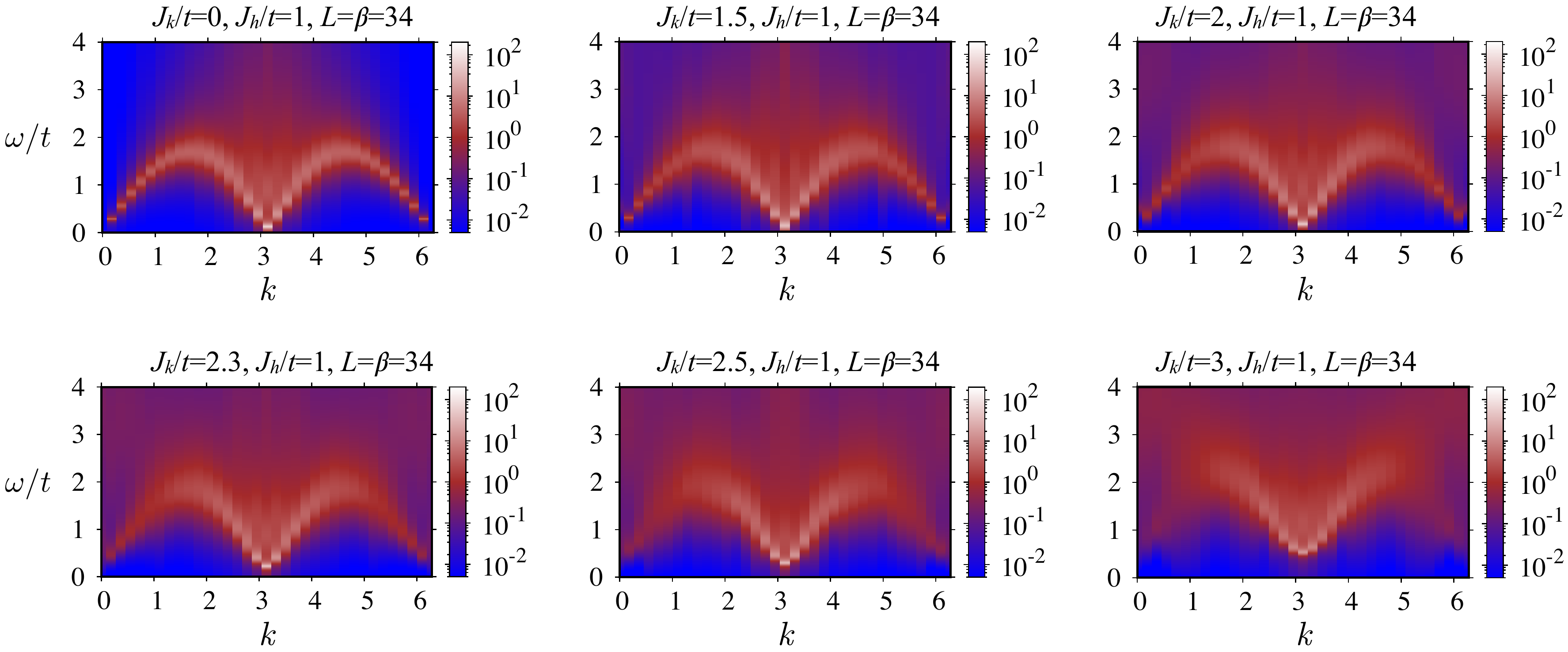}
\caption{Dynamical spin structure factor, $S(k,\omega)$, as a function of  energy ($\omega/t$) and momentum ($k$) along the spin chain for various values of $J_k/t$ on $\beta=L=34$ size systems.}
\label{Skomega_vs_omega_4n+2_all}

~

\includegraphics[width=0.96\textwidth]{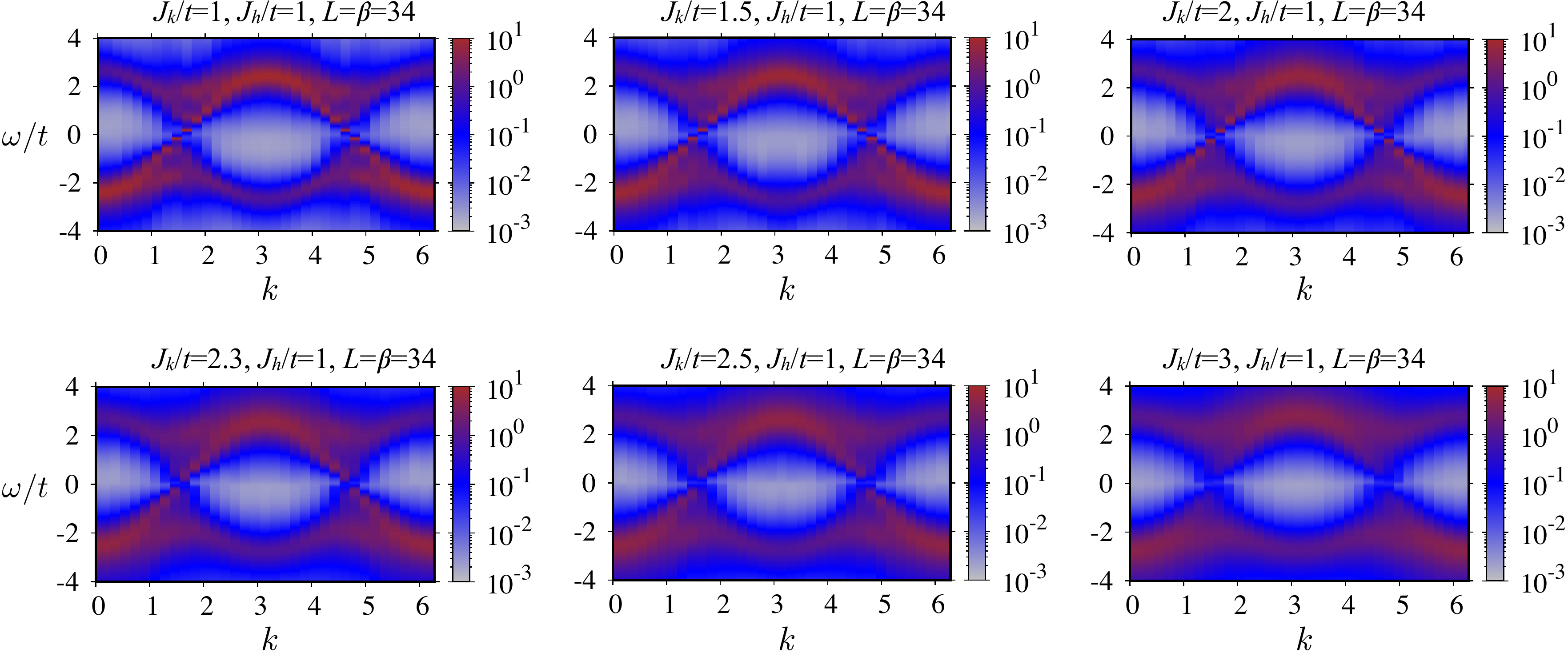}
\caption{Single particle excitations, $A_0(k,\omega)$, as a function of  energy ($\omega/t$) and momentum ($k$) along Kondo coupled row of conduction electrons for various values of $J_k/t$ on $\beta=L=34$ size systems.}
\label{Ckomega_vs_omega_4n+2_all}
\end{figure}
\end{document}